\documentclass[twocolumn,english,aps,prl,floatfix,amssymb,superscriptaddress,longbibliography]{revtex4}
\usepackage[latin9]{inputenc}
\usepackage{verbatim}
\usepackage{float}
\usepackage{amsmath}
\usepackage{amssymb}
\usepackage{graphicx}
\usepackage{color}
\usepackage{xcolor}
\usepackage{tikz}
\newcommand{\ket}[1]{\ensuremath{\left| #1 \right>}}

\newcommand{\Tr}{\text{Tr}}
\usepackage{bbold}

\usepackage[bookmarks=false,linkcolor=blue,urlcolor=blue,colorlinks,citecolor=blue]{hyperref}
\makeatletter

\newcommand{\be}{\begin{equation}}
\newcommand{\ee}{\end{equation}}
\newcommand{\bea}{\begin{eqnarray}}
\newcommand{\eea}{\end{eqnarray}}

\begin{document}

\title{Boundary-Induced Topological and Mid-Gap States in Charge Conserving One-Dimensional Superconductors}
\author{Parameshwar R. Pasnoori}
\affiliation{Department of Physics and Astronomy, Rutgers University, Piscataway, NJ 08854-8019 USA}
\author{Natan Andrei}
\affiliation{Department of Physics and Astronomy, Rutgers University, Piscataway, NJ 08854-8019 USA}
\author{Patrick Azaria}
\affiliation{Laboratoire de Physique Th\'eorique de la Mati\`ere Condens\'ee, Sorbonne Universit\'e and CNRS, 4 Place Jussieu, 75252 Paris, France}

\begin{abstract}
We investigate one-dimensional charge conserving, spin-singlet (SSS) and spin-triplet (STS) superconductors in the presence of boundary fields. In systems with Open Boundary Conditions (OBC) it has been demonstrated that STS display a four-fold topological degeneracy, protected by
the $\mathbb{Z}_2$ symmetry which reverses the spins of all fermions,
whereas SSS are topologically trivial.  In this work we show  that it is not only the type of the bulk superconducting instability that determines the eventual topological nature of a phase, but rather the interplay between bulk and boundary properties.  In particular we show by means of the Bethe Ansatz technique that SSS may as well be in a $\mathbb{Z}_2$-protected topological phase provided suitable "twisted" open boundary conditions ${\widehat{OBC}}$ are imposed. More generally, we find that depending on the boundary fields, a given superconductor, either SSS or STS,  may exhibits several types of phases such as topological, mid-gap  and trivial phases; each phase being characterized by a boundary fixed point which  which we determine. Of particular interest are the mid-gap phases which are stabilized close to the topological fixed point. They include both fractionalized
 phases where spin-$\frac{1}{4}$ bound-states are localized at the two edges of the system 
 and un-fractionalized phases where a spin-$\frac{1}{2}$ 
 bound-state is localized at either the left or the right
 edge.

\end{abstract}
\maketitle

\section{Introduction.}
Symmetry Protected Topological (SPT) phases of matter, as their name suggest, display gapless end modes whose stability require symmetry protection\cite{Wen, Starykh2000, Turner2011,Sau2011,Fidkowski2011b, Beenakker2013, ruhman2015, ruhman2017topological, jiang2017symmetry, kainaris2017interaction, Keselman2018, PAA,colinmidgap}.  
Prototypical examples of   such systems are    one-dimensional charge conserving Spin-Triplet Superconductor (STS)\cite{Keselman2015} which  exhibit  two zero-energy Majorana (ZEM) modes  at each end of an open chain, whose protection  is insured by the $\mathbb{Z}_2$ symmetry consisting of flipping the spins of all the fermions. Due to the presence of the four ZEM,
the ground-state degeneracy is four-fold and is exhausted by fractional spin states $\pm 1/4$ exponentially localized at the ends of the system. Spin-Singlet Superconductors (SSS) on the other hand are topologically trivial: they do not display gapless Majorana  modes in an open system and their ground-states  is unique with total spin $S^z=0$.
The two STS and SSS phases  cannot be connected adiabatically, while maintaining the $\mathbb{Z}_2$ symmetry,  without closing the gap in the bulk.
 Therefore in  1-D superconductors   the nature of the  {\it bulk} superconducting  instability  is intrinsically linked to the topological nature of the phase. 
 
 Though this is certainly true as far as  Open Boundary Conditions (OBC) are considered, we shall demonstrate in the present work  that in the presence of {\it boundary} fields  the situation changes drastically. We will show that it is not only the bulk superconducting instabilities  that determine the topological nature of a phase but rather the interplay between  bulk and boundary properties. In the following,  we shall solve exactly, by the Bethe Ansatz (BA) technique,  the continuum Hamiltonian relevant for $1D$ charge conserving superconductors with  {\it arbitrary} integrable $\mathbb{Z}_2$ symmetry-breaking boundary conditions. As the resulting phase diagram is rich and complex we shall summarize our main results before going into details. 
 
 \section{Main results} One of our main finding is that  in one-dimensional superconductors  it is not  only the type of the bulk superconducting instabilities that determine the topological nature of a phase but rather the interplay between  bulk and boundary properties.  In particular it will be shown that, when appropriate boundary conditions,  to be referred to as  "twisted" boundary conditions $\widehat{OBC}$,  are applied at the two ends of an open chain, the seemingly trivial $SSS$ phase becomes topological in that  it exhibits two protected zero-energy Majorana  modes at each end of the system.  At the same time  the $STS$ with twisted $\widehat{OBC}$ is rendered topologically trivial and does not  exhibit ZEM. Hence, both $STS$ and $SSS$ can exhibit protected ZEM depending on whether one applies $OBC$ or $\widehat{OBC}$ boundary conditions respectively. 
\begin{center}
\begin{figure}[!h]
\includegraphics[width=0.6\columnwidth]{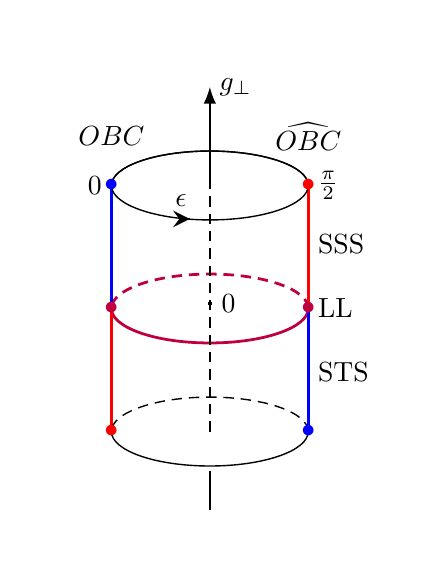}
\caption{Qualitative phase diagram of charge conserving superconductors in the presence of a boundary field.
The coupling $g_{\perp}$ controls the interaction in the bulk and $\epsilon$ is a twist angle parametrizing the boundary conditions. In the $U(1)$-Thirring model under study this corresponds to $g_{\parallel} > 0$ in (\ref{Hamiltonian}) and $\epsilon_{\cal{R}}=\epsilon_{\cal{L}}=\epsilon$ for the boundary conditions (\ref{IntBC}).
A bulk SSS (STS) instability corresponds to $g_{\perp}>0$ ($g_{\perp}<0)$.  $\mathbb{Z}_2$ symmetric OBC ($\widehat{OBC}$) corresponds to $\epsilon=0$ ($\epsilon=\pi/2$). For a fixed boundary condition twist $\epsilon=(0,\pi/2)$, the system undergoes a quantum phase transition from a trivial
 phase (blue line) to a topological phase (red line)
 at $g_{\perp}=0$ which is  Luttinger Liquid (LL) phase (dark red line). Fixing $g_{\perp} >0$  $(g_{\perp} <0)$  as one varies the boundary twist $\epsilon$, the
 $\mathbb{Z}_2$ symmetry is broken, and one may go 
 from a trivial (topological) phase when $\epsilon =0$ ($\epsilon =\pi/2$) to a topological (trivial) phase
 at $\epsilon =\pi/2$ ($\epsilon =0$). In the process, before  reaching  the topological phase, one enters a mid-gap region which  physics is controlled by a topological fixed point.
}
\label{fig:1}
\end{figure}
\end{center}
As we shall see, the two pairs of bulk instability-boundary condition $SSS$-$\widehat{OBC}$ and $STS$-$OBC$ are stabilized by  a single  $\mathbb{Z}_2$ symmetric topological boundary fixed point while the pairs  $SSS$-$OBC$ and $STS$-$\widehat{OBC}$ correspond to a trivial $\mathbb{Z}_2$ symmetric fixed point that controls the trivial phase. It is only close to these two fixed points  that  universal results can be obtained from the BA solution in the scaling limit. 
One may move from a topological phase (either $SSS$ with $\widehat{OBC}$ or $STS$ with  $OBC$)  to a trivial  one  (either $SSS$ with $OBC$ or $STS$
with $\widehat{OBC}$) {\it without} going through a quantum phase transition in the bulk,  only changing continuously the boundary conditions from $OBC$ to  $\widehat{OBC}$ or vice-versa (see Fig\ref{fig:1}). In so doing, one inevitably breaks the  $\mathbb{Z}_2$ symmetry during the path (see Fig. \ref{fig:1}). As a result,  close to the topological fixed point, the Majorana end modes become gapped and   turn into {\it mid-gap} states. The whole  mid-gap region, which is stabilized by the topological fixed point, precludes the topological degeneracy point.
We shall argue that, close enough to the topological fixed point,  these mid-gap states share the same quantum numbers and span  the same  Hilbert space  as the ZEM: they are localized modes which span fractionalized representations of the symmetry groups of the Hamiltonian.  This is quite fortunate since in an open system  the environment acts on the boundaries in all possible ways which are generically  not $\mathbb{Z}_2$ symmetric. 
As one departs farther from the topological fixed point,
some of the Majorana end-modes leak into the bulk and fractionalization is lost. Despite this there still
exists  mid-gap states which correspond to localized
spin-$\pm \frac{1}{2}$ bound-states localized at either
the left or the right edge.
When one moves too far away from the topological fixed  point, the mid-gap states eventually disappear from the spectrum and leak into the bulk.  Finally,  close to  the trivial fixed point  the system displays a non degenerate singlet ($S^z=0$) ground-state and is in a topologically trivial state with universal properties.  In the region in between the two topological and trivial fixed points  the nature of a possible boundary phase transition between topological and trivial phases  remains an open question.

 \section{1D superconductors with boundaries.}The Hamiltonian we shall consider is that  of the $U(1)$ Thirring model  given by $H= \int_{-L/2}^{L/2} dx\;   {\cal H}(x)$ where, 
\bea \label{Hamiltonian}
 {\cal H}&=&  -i v \left(  \psi^{\dagger}_{Ra} \partial_x \psi_{Ra} -  \psi^{\dagger}_{La} \partial_x \psi_{La}\right)
 \\
&+&
   \psi^{\dagger}_{Ra} \psi_{Rb}  [  \,g_{\parallel}\;  \sigma^z_{ab} \sigma^z_{cd} + g_{\perp}\; (\sigma^x_{ab} \sigma^x_{cd}+ \sigma^y_{ab} \sigma^y_{cd})] \psi^{\dagger}_{Lc}\psi_{Ld},\nonumber 
\eea
where  $\psi_{Ra}, \psi_{La}$, $a=(\uparrow, \downarrow)$,  are   two-component spinor fields   describing  right (R) and  left (L) moving spin-$1/2$ fermions and  $\sigma^{x,y,z}$ are the Pauli matrices. The Hamiltonian (\ref{Hamiltonian}) has been shown to be integrable with periodic boundary conditions \cite{Andrei,duty,Japaridze} and with $OBC$ in the STS phase \cite{PAA}. The Hamiltonian (\ref{Hamiltonian}) is invariant under $U(1)_c$
and $U(1)_s$ symmetries, in charge and spin sectors respectively,  with associate conserved charges, $N={\textstyle \int_{-L/2}^{L/2} dx\;[\psi^{\dagger}_{Ra} \psi_{Ra} +  \psi^{\dagger}_{La}  \psi_{La}}]$, and spin, $S^z={\textstyle \int_{-L/2}^{L/2} dx\;[\psi^{\dagger}_{Ra} \sigma^z_{ab}\psi_{Rb} +  \psi^{\dagger}_{La} \sigma^z_{ab} \psi_{Lb}}]$. On top of the above, (\ref{Hamiltonian}) is also invariant under the $\mathbb{Z}_2$ symmetry which exchanges the spins of the fermions
\be
\psi_{R(L)} \rightarrow \sigma^x \psi_{R(L)}, 
\label{z2}
\ee
and reverses the total spin $S^z \rightarrow -S^z$ of the system.

We shall impose the following boundary conditions at the  left and the right ends of the system, i.e: at $x=\mp L/2$
\bea
\psi_{R}(-L/2)&=& - B^{\cal{L}}(g)\;   \psi_{L}(-L/2), \\ \nonumber 
  \psi_{R}(+L/2)&=& - B^{\cal{R}}(g)\;   \psi_{L}(+L/2),
  \label{BC}
\eea
where $B^{\cal{L},\cal{R}}(g)$ are $2 \times 2$ diagonal matrices acting on the spin components  of the spinors and  which may depend on the couplings $g\equiv(g_{\parallel}, g_{\perp})$. 
For OBC, namely  when $B^{\cal{L}}=B^{\cal{R}} = I$, the system is known to exhibit the two superconducting phases $SSS$ and $STS$ in the domains
 $g_{\parallel} >  g_{\perp} > 0$ and  $g_{\parallel} >  -g_{\perp} > 0$ respectively.
As already mentioned, in the latter domain of couplings the system exhibits four protected 
Majorana ZEM localized at the ends of the chain. We shall now see that this 
implies that the $SSS$ phase may become topological when suitable boundary conditions are imposed.

\subsection{Duality and Twisted OBC.} The reason  for the last statement  stems from a hidden duality symmetry $\Omega$ \cite{BoulatDuality} of the  Hamiltonian
\be
H(\psi, g_{\parallel},  g_{\perp}, B^{\cal{L}}, B^{\cal{R}}) = H(\widehat \psi, g_{\parallel},  -g_{\perp}, \widehat{B^{\cal{L}}}, \widehat{B^{\cal{R}}}),
\label{dualhamiltonian}
\ee
where the duality $\Omega$ acts on the fermions as $\widehat \psi = \Omega  \psi$ with 
\bea
\widehat \psi^{}_{L}= \psi^{}_{L}, \; \widehat \psi^{}_{R} = i  \sigma^z  \psi^{}_{R},
\label{duality}
\eea
and on the boundary conditions as
\be
\widehat{B^{\cal{L},\cal{R}} }=  i  \sigma^z B^{\cal{L},\cal{R} }(-g_{\perp}).
\label{dualBC}
\ee
We immediately see that $\Omega$ maps the $STS$ phase, with $g_{\perp} < 0$ and   $OBC$  i.e. $B^{\cal{L},\cal{R} }=I$, to an $SSS$ phase with $g_{\perp} > 0$ and {\it twisted} 
$\widehat{OBC}$ with $\widehat{B^{\cal{L},\cal{R} }}=  i  \sigma^z$. Since from (\ref{dualhamiltonian})
both systems have the same spectrum, we deduce that a $SSS$ with twisted $\widehat{OBC}$ displays four zero energy Majorana modes localized at the ends of the system, exactly as for the topological $STS$ with $OBC$. The topological degeneracy in this case is boundary induced and  is still protected by the $\mathbb{Z}_2$ symmetry (\ref{z2}).
Indeed, although under $\mathbb{Z}_2: B^{\cal{L},\cal{R}} \rightarrow - B^{\cal{L},\cal{R}}$, both
boundary conditions $\pm B^{\cal{L},\cal{R}} $ are equivalent since the Hamiltonian is invariant under the
independent changes $\psi^{}_{L,R} \rightarrow -\psi^{}_{L,R}$. Similarly choosing $B^{\cal{L},\cal{R} }= i  \sigma^z$ one finds that  $STS$ are topologically trivial when twisted $\widehat{OBC}$ are considered. 
In both $SSS$-$\widehat{OBC}$ and $STS$-$OBC$ systems,  the ground-state displays a four-fold degeneracy.  The four ground states are  labelled by their total spins $S^z=( \pm \frac{1}{2},0)$:
\be
\{|-\frac{1}{2}\rangle, |0\rangle, |0\rangle^{'}, |+\frac{1}{2}\rangle\},
\label{SSStwistedgs}
\ee
and transform into each other under the $\mathbb{Z}_2$ symmetry generator $\sigma^x$(\ref{z2}). Moreover, as discussed in \cite{Wen,Keselman2018,PAA}, the system  support fractional spin states 
$|\pm \frac{1}{4}\rangle_{\cal{L},\cal{R}}$, localized at each edge of the system, with local spins
$S^z_{\cal{L},\cal{R}}= \frac{1}{4} \sigma^z_{\cal{L},\cal{R}}$ such that
$S^z=S^z_{\cal{L}}+S^z_{\cal{R}}$ and
\bea
\label{fracspin}
|\pm \frac{1}{2}\rangle &=& |\pm \frac{1}{4}\rangle_{\cal{L}} \otimes |\pm \frac{1}{4}\rangle_{\cal{R}},\nonumber \\
|0\rangle &=& |-\frac{1}{4}\rangle_{\cal{L}} \otimes |+\frac{1}{4}\rangle_{\cal{R}}, \nonumber \\
|0\rangle^{'} &=& |+\frac{1}{4}\rangle_{\cal{L}} \otimes |-\frac{1}{4}\rangle_{\cal{R}}. 
\eea
The states  $|\pm \frac{1}{4}\rangle_{\cal{L},\cal{R}}$  span  a representation of the fractionalized $\mathbb{Z}_2$ symmetry (\ref{z2}), i.e: $\mathbb{Z}_2= \mathbb{Z}_{2\cal{L}} \times \mathbb{Z}_{2\cal{R}}$, where  $\mathbb{Z}_{2\cal{L},\cal{R}}$ 
are generated by  two Majorana fermions  localized at the left and right boundaries, $\sigma^x_{\cal{L}}$ and $\sigma^x_{\cal{R}}$, such that  $\sigma^x_{\cal{L},\cal{R}}|\pm \frac{1}{4}\rangle_{\cal{L},\cal{R}}=|\mp \frac{1}{4}\rangle_{\cal{L},\cal{R}}$. Thus, each edge support two Majorana fermions, $(\sigma^x_{\cal{L},\cal{R}}, \sigma^y_{\cal{L},\cal{R}})$,
$\sigma^y_{\cal{L},\cal{R}} =- i \sigma^x_{\cal{L},\cal{R}}\sigma^z_{\cal{L},\cal{R}}$, which act
in the Hilbert space of fractional spin states. These Majorana modes are the zero-energy modes, in the thermodynamical limit, which characterize the topological state in the $SSS$-$\widehat{OBC}$ and $STS$-$OBC$ systems.

The question we shall  now address is whether the stability of  topological and the trivial phases when the $\mathbb{Z}_2$ symmetry is broken by considering small twists around both  $OBC$ and  $\widehat{OBC}$.
To answer this  question we shall solve exactly, by means the Bethe Ansatz, the  $SSS$  Hamiltonian (i.e: $g_{\perp} > 0$) with  the  most general integrable boundary conditions, which are  diagonal in spin space. Results for the  $STS$ Hamiltonian (i.e: $g_{\perp} < 0$) can  be obtained using the  duality symmetry (\ref{duality}).

 \section{Stability of the  $OBC$ and twisted  $\widehat{OBC}$: the Bethe anstaz solution.} 
Integrable boundary conditions correspond to specific choices of the boundary matrices 
(\ref{BC}) that satisfy the Boundary Yang-Baxter (BYB) equations \cite{Sklyannin,Cherednik}. We find that, up to a phase $B^{\cal{L},\cal{R}} \rightarrow  e^{\Phi_{\cal{L},\cal{R}}}B^{\cal{L},\cal{R}}$ ($\Phi_{\cal{L},\cal{R}} \in \mathbb{C}$), the most general diagonal integrable boundary conditions are given by the matrices \footnote{From the structure of these equations one may show that if $B^{L,R}$ are solutions of the BYB then $e^{\Phi_{L,R}}B^{L,R}$, where $\Phi_{L,R}$ are arbitrary (complex) phases,  are also solutions. The only effect of the phases $\Phi_{L,R}$ is to shift the energies of all the eigenstates of (\ref{Hamiltonian}) by the same amount so that $B^{L,R}$ and  $e^{\Phi_{L,R}}B^{L,R}$ yield to equivalent solutions.},

\bea
\label{IntBC}
B^{\cal{R}}&=& \frac{1}{\cosh( \displaystyle{\frac{f}{2}})}\left(\begin{array}{cc} \cosh( \displaystyle{\frac{f}{2}}+i\epsilon_{\cal{R}})&0\\0&\cosh( \displaystyle{\frac{f}{2}}-i\epsilon_{\cal{R}})\end{array}\right),
\nonumber \\
B^{\cal{L}}&=&\frac{1}{\cosh( \displaystyle{\frac{f}{2}})}\left(\begin{array}{cc} \cosh( \displaystyle{\frac{f-iu}{2}} +i\epsilon_{\cal{L}})&0\\0&\cosh( \displaystyle{\frac{f-iu}{2}}-i\epsilon_{\cal{L}})\end{array}\right),
\nonumber \\
\eea
where the parameters $(f,u)$ are related to the couplings $(g_{\parallel}, g_{\perp})$ entering in (\ref{Hamiltonian}) by \cite{duty}
  \bea\cos(u)=\frac{\cos(g_{\parallel})}{\cos(g_{\perp})}, \hspace{2mm} \frac{\sin(u)}{\tanh(f)}=\frac{\sin(g_{\parallel})}{\cos(g_{\perp})}.\eea
The angles $\epsilon_{\cal{L}}$ and $\epsilon_{\cal{R}}$,  $\epsilon_{{\cal{L}},{\cal{R}}} \in [-\pi/2, \pi/2]$, are independent twists parametrizing  the boundary conditions at the left and right ends of the system. Physically, the above boundary conditions on the fermions can be seen as the effect of applying a magnetic field along the $z$-axis which is localized around the left and right boundaries.
$OBC$ and twisted $\widehat{OBC}$ are obtained with the twists
$\epsilon_{\cal{L}}=\epsilon_{\cal{R}}=0$ and $\epsilon_{\cal{L}}=\epsilon_{\cal{R}}= \pi/2$ 
for which  $B^{{\cal{L}},\cal{R}}\propto I$ and $B^{\cal{L},\cal{R}}\propto i \sigma^z$ respectively.
Under the  $\mathbb{Z}_2$ symmetry of eqn(\ref{z2})  we have
$B^{\cal{L},\cal{R}} \rightarrow \sigma^x B^{\cal{L},\cal{R}} \sigma^x$
or equivalently
\be 
\mathbb{Z}_2: \epsilon_{{\cal{L}},{\cal{R}}} \rightarrow  -\epsilon_{{\cal{L}},{\cal{R}}}.
\label{z2epsilon}
 \ee
Hence, despite the fact that the bulk Hamiltonian is invariant under $\mathbb{Z}_2$, generic boundary conditions
break the $\mathbb{Z}_2$ symmetry.
The only invariant boundary conditions are $OBC$ and twisted $\widehat{OBC}$.

\smallskip

We have obtained the complete solution of the Hamiltonian (\ref{Hamiltonian}) with the boundary conditions (\ref{IntBC}) using the Boundary Algebraic Bethe Ansatz. The resulting Bethe equations, as well as their derivation,  are given in the Appendix. We shall present in what follows the ground-states properties as well as that of the low-energy excitations in the different phases of the problem obtained when one varies the twists $\epsilon_{\cal{L},\cal{R}}$ while keeping fixed the bulk couplings, $g_{\parallel, \perp}$.

\paragraph{Scaling limit.} There are four regimes of twists where universal results can be obtained in the scaling limit. Each regime is characterized by a fixed point, and related RG invariants, which we now define.

Independently of the boundary conditions, the bulk physics is characterized by the opening of a single particle gap \cite{Japaridze,Andrei}
\be
m= D \; \{\arctan[\sinh(\frac{\pi f}{2u})]\}^{-1}
\label{gap}
\ee
where $D=N/L$ is an ultra-violet cut-off, $N$ being the total number of fermions and $L$
the size of the system. Universality is obtained in the limit $D\rightarrow \infty$ and $u\rightarrow 0$ while keeping the physical mass $m$ fixed. 
This corresponds to the weak coupling regime $g_{\parallel, \perp} \ll 1$ where quantum fluctuations are strong. 
From the Bethe equations we find that this implies a scaling limit
on the twist angles $\epsilon_{\cal{L},\cal{R}}$ that have to scale to  $\epsilon^{*}_{\cal{L},\cal{R}}=0$ or $\epsilon^{*}_{\cal{L},\cal{R}}=\pi/2$.

In the following,  we shall investigate in details the effects of the boundary
twists close to both $OBC$ and $\widehat{OBC}$. Both regions are governed by two fixed points that we shall call trivial and topological fixed
points at $(\epsilon^*_{\cal{L}},\epsilon^*_{ \cal{R}}) = (0,0)$ and $(\epsilon^*_{\cal{L}},\epsilon^*_{ \cal{R}}) = (\pi/2,\pi/2)$.
This defines, close to each of these two fixed points, RG-invariant parameters 
\be
\label{epsilonprime}
\epsilon^{'}_{L,R}= \frac{\epsilon_{\cal{L},\cal{R}} - \epsilon^{*}_{\cal{L},\cal{R}}}{2u} ,
\ee
where $\epsilon^*_{\cal{L}}=\epsilon^*_{ \cal{R}}=0$ and $\epsilon^*_{\cal{L}}=\epsilon^*_{ \cal{R}}=\pi/2$,
 that are kept fixed in the scaling limit: $ \epsilon_{\cal{L},\cal{R}} \rightarrow \epsilon^{*}_{\cal{L},\cal{R}}, u\rightarrow 0$.
The $\epsilon^{'}_{\cal{L},\cal{R}}$ are the physical twists parameters which, together with the  mass  $m$ in (\ref{gap}),  determine the universal physical properties of the system. 
The two fixed points, which are associated with specific boundary conditions, stabilize two different scaling regions that we shall describe in the following.

 \subsection{Trivial Region\label{trivialR}} This is the region which corresponds  to the $OBC$ fixed point $(\epsilon^{*}_{\cal{L}}, \epsilon^{*}_{\cal{R}})=(0,0)$. 
  In the hole domain the ground-state is a singlet with
  $S^z  = 0$ and is non degenerate. In this region, the boundaries
  play a minor role and the physics is qualitatively similar to what happens with periodic boundary conditions.

 \subsection{Topological Region\label{topologicalR}}
 \begin{center}
\begin{figure}[!h]
\includegraphics[width=1.1\columnwidth]{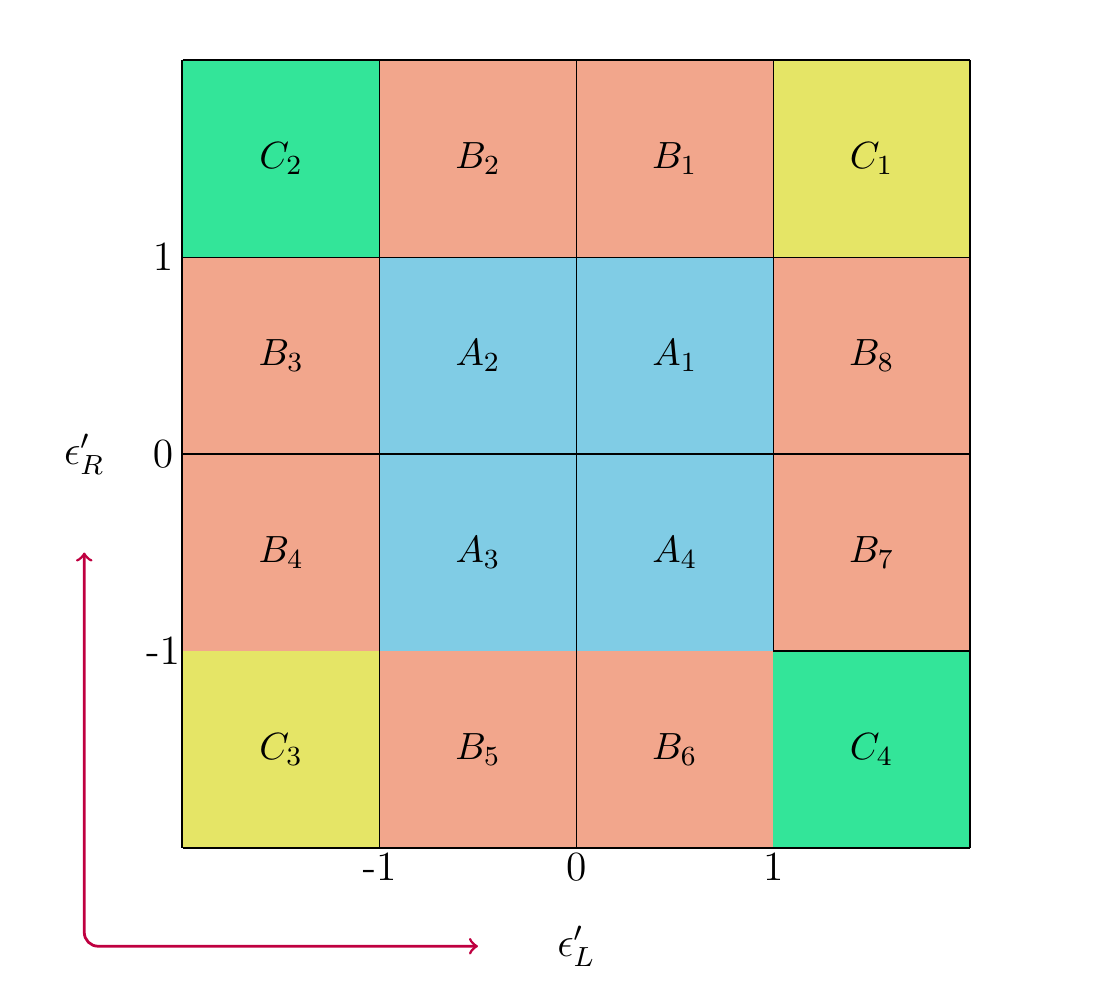}
\caption{\label{fig:2} The figure shows various phases corresponding to the topological region. There are four low lying states in the blue region representing phases $A_j$, and they form a representation of $\mathbb{Z}_2$. There are two low lying states in the red region which represents phases $B_j$, and there is only one low lying state which is the ground state in the yellow and green regions representing phases $C_j$. Tables \ref{table1}, \ref{table2} and \ref{table3} summarize the energies of low lying states in the phases $A_j$, $B_j$ and $C_j$ respectively. Fractionalization-non fractionalization phase transition occurs at the boundary between $A_j$ and $B_j$ or $C_j$. At boundaries within the blue region and within the red regions, as the signs of $\epsilon'_i$ change, level crossings occur between the low lying states leading to first order phase transitions.}
\end{figure}
\end{center}
 This is the  region which is  stabilized by   the twisted $\widehat{OBC}$   fixed point at $(\epsilon^{*}_{\cal{L}}, \epsilon^{*}_{\cal{R}})=(\pi/2,\pi/2)$. This region is characterized
 by the existence of mid-gap states which are reminiscent of boundary states localized at the left and/or right
 boundaries. The  topological region further splits into three regions A, B and C   depending on  the number of mid-gap states ${\cal N}=(3,1,0)$ respectively. These sub-regions extend in different domains
 of the  twists   $\epsilon^{'}_{\cal{L},\cal{R}}$.
 
\subsubsection{Region A\label{regionA}}

When $|\epsilon^{'}_{\cal{L},\cal{R}}| < 1$ the number of low-energy states and their spins are the same that  at the topological fixed point $SSS$-$\widehat{OBC}$ (\ref{SSStwistedgs}). Everywhere in region A there exist  four  low-lying  states, with spins $S^z= \pm1/2$ and $S^z=0$,
 \be
 \label{Astates}
\{  |-\frac{1}{2} \rangle, |0\rangle_{\epsilon^{'}_{\cal{L}}}, |0\rangle_{\epsilon^{'}_{\cal{R}}}, |+\frac{1}{2}\rangle\},
 \ee
  which have  fermion parities  ${\cal P}= (-1)^N= e^{i 2\pi S^z} = (-1, +1, +1, -1)$ respectively. 
   In the limit $\epsilon^{'}_{\cal{L},\cal{R}} \rightarrow  0$ they identify   with the four degenerate states (\ref{SSStwistedgs}) 
  at the topological fixed point with the correspondence $|0\rangle_{\epsilon^{'}_{\cal{L}}} \rightarrow |0'\rangle$
  and $|0\rangle_{\epsilon^{'}_{\cal{R}}} \rightarrow |0\rangle$.
  When $\epsilon^{'}_{\cal{L},\cal{R}} \neq 0$ 
 the four-fold degeneracy is lifted:  some of the spin states  (\ref{Astates}) become {\it mid-gap} states.
In the Bethe Ansatz approach, in the phase $A_1$, the states $|+\frac{1}{2} \rangle$ and $|-\frac{1}{2} \rangle$  are constructed from Bethe reference states with all spin up and all spin down respectively, and they both have all real Bethe roots. The two singlet states $ |0\rangle_{\epsilon^{'}_{\cal{L}}}$ and $ |0\rangle_{\epsilon^{'}_{\cal{R}}}$ are obtained by adding  imaginary boundary strings solutions of the Bethe equations (which carry a spin $-\frac{1}{2}$), $\lambda_{\epsilon^{'}_{\cal{R}}}= \pm i (1-\epsilon^{'}_{\cal{R}})/2$  and $\lambda_{\epsilon^{'}_{\cal{L}}}= \pm i (1-\epsilon^{'}_{\cal{L}})/2$ respectively,
to  the state $|+\frac{1}{2}\rangle$.  These solutions correspond to boundary bound-states \cite{skorik,mez}, localized at the left and right boundaries of the system,  with energies $- m_{\cal{L}}$ and $-m_{\cal{R}}$ relative to that of the $|+\frac{1}{2}\rangle$ state where, in the scaling limit,
\be\label{bsenergy}
 m_{\cal{L}, \cal{R}} = m\; \sin{\left(\frac{\pi}{2} \epsilon^{'}_{\cal{L},\cal{R}}\right)}, \; |\epsilon^{'}_{\cal{L},\cal{R}}| < 1.
 \ee
On the other hand, the energy difference
$\Delta E$ between the states $|+\frac{1}{2}\rangle$ and $|-\frac{1}{2} \rangle$ can be shown to be equal
to $\Delta E= m_{\cal{L}}+m_{\cal{R}}$. Measuring all energies with respect 
 to that of the $S^z= -1/2$ state, we get for the energies of the four states (\ref{Astates}) (see table (\ref{table1}))
\bea\label{midgap}
 E_{-\frac{1}{2}},  \; E_{-\frac{1}{2}}+m_{\cal{L}}, \;E_{-\frac{1}{2}} + m_{\cal{R}},\;  E_{-\frac{1}{2}}+  m_{\cal{L}} + m_{\cal{R}} 
\eea

 In the remaining $A_{2,3,4}$ phases, as discussed in the Appendix,  the  spin quantum numbers and corresponding  energies  are the same as is the $A_1$ phase although the construction of the four low-energy states is different. The resulting  arrangements of the four states in each phase eventually depend  on the signs of $\epsilon^{'}_{\cal{L},\cal{R}}$, and hence on those  of the mid-gap energies $m_{\cal{L},\cal{R}}$. We further
 distinguish between four phases $A_{1}, A_{2}, A_{3}$ and  $A_{4}$ depending on which of the four states (\ref{Astates}) is the ground-state. In these phases, which  are stabilized  in the following domains of twists, $A_{1}:( \epsilon^{'}_{\cal{L}} > 0, \epsilon^{'}_{\cal{R}} >0)$, $A_{2}:( \epsilon^{'}_{\cal{L}} <0 ,  \epsilon^{'}_{\cal{R}} >0)$,  $A_{3}:( \epsilon^{'}_{\cal{L}} <0 ,  \epsilon^{'}_{\cal{R}} <0)$, and $A_{4}:( \epsilon^{'}_{\cal{L}} > 0,  \epsilon^{'}_{\cal{R}} <0)$, 
 the ground-state is in a different spin state, i.e:  $|-\frac{1}{2} \rangle,  |+\frac{1}{2}\rangle,  |0\rangle_{\epsilon^{'}_{\cal{L}}}$ or $ |0\rangle_{\epsilon^{'}_{\cal{R}}}$ respectively.
The remaining three states further order, in each phase, according  to the relative values of $m_{\cal{L},\cal{R}}$. We summarize our results in the Table (\ref{table1}). 

 \begin{table}[h!]
\centering
\caption{Total spin and energy in the scaling limit of low-lying states in the  phase $A_1$. The energies are measured with respect to $E_{-\frac{1}{2}}$ and $m_{ \cal{L}}, m_{ \cal{R}}$ are given by Eq.(\ref{bsenergy}).}

\begin{tabular}{ccc}
\hline
\hline
  State  & \;\;Total spin &\;\; Energy \\
\hline
 $|-\frac{1}{2} \rangle$& -1/2  & 0    \\
  $|0\rangle_{\epsilon^{'}_{\cal{L}}}$ & 0       &   $m_{\cal{R}}$    \\
    $|0\rangle_{\epsilon^{'}_{\cal{R}}} $ & 0  & $m_{\cal{L}}$  \\
$|+\frac{1}{2}\rangle$ & 1/2 & $m_{\cal{L}}+m_{\cal{R}}$\\
\hline
\hline
\end{tabular}
\label{table1}
\end{table}  
The four phases  $A_{j=1,...,4}$  transform into each other under the $\mathbb{Z}_2$ group generator  (\ref{z2epsilon})
as $A_1 \leftrightarrow A_3$ and $A_2 \leftrightarrow A_4$ and are invariant under space parity which exchanges  the two boundaries 
$\mathbb{P}:\cal{L} \leftrightarrow \cal{R}$. Remarkably  enough,
we stress that although, when 
$\epsilon^{'}_{\cal{L},\cal{R}}\neq 0$, the boundary conditions break the $\mathbb{Z}_2$ symmetry, the four states in {\it each} phase $A_{j=1,...,4}$ transform  into each other under  $\mathbb{Z}_2$ and  hence span a representation of the $\mathbb{Z}_2$  group $S^z \rightarrow -S^z$. This fact has important consequences
on the nature of the low-energy Hilbert space spanned
by the four states (\ref{Astates}) as we shall discuss in the next section. For the time being, we observe 
that  the four phases  $A_{j=1,...,4}$ are separated by   boundary quantum phase transitions lines at $\epsilon^{'}_{\cal{L}}=0$ or $ \epsilon^{'}_{\cal{R}}=0$ where one of the mid-gap energies $m_{\cal{L}}$ or $m_{ \cal{R}} $ closes. The phase transitions lines between $A_{1} \leftrightarrow A_{2}$ and $A_{3} \leftrightarrow A_{4}$ is  at  $ \epsilon^{'}_{\cal{L}} = 0$  for   $\epsilon^{'}_{\cal{R}} >0$ and $\epsilon^{'}_{\cal{R}} <0$, whereas that between $A_{1} \leftrightarrow A_{4}$ and $A_{2} \leftrightarrow A_{3}$ is  at 
$ \epsilon^{'}_{\cal{R}} = 0$  for   $\epsilon^{'}_{\cal{L}} >0$ and $\epsilon^{'}_{\cal{L}} <0$.
When crossing these lines, as either $m_{\cal{L}}$ or $m_{ \cal{R}} $ changes its  sign, there are level crossings between pairs of states with opposite fermion parities. At the phase transition point, the ground-state is doubly degenerated with the two ground-states having opposite fermion parities ${\cal P}= \pm 1$. As a consequence,
when $ \epsilon^{'}_{\cal{L}, \cal{R}} = 0$,
there is a zero energy mode (in the thermodynamical limit) corresponding to adding a fermion, with either spin $\uparrow$ or $\downarrow$,   at either the left or the right boundary. When $ \epsilon^{'}_{\cal{L}} =   \epsilon^{'}_{ \cal{R}} = 0$, i.e. at the topological point,  there two of them \cite{PAA}.

Consider for instance the phase transition between the sub-regions $A_{1}$ and $ A_{2}$ for which $m_{ \cal{R}}  > 0$. When going from $A_{1}$ to $ A_{2}$,  by varying $ \epsilon^{'}_{\cal{L}}$ from positive values to negative values, the two states in each pair of states ($|-\frac{1}{2} \rangle$,  $|0\rangle_{\epsilon^{'}_{\cal{R}}}$) and ( $|0\rangle_{\epsilon^{'}_{\cal{L}}}$, $|\frac{1}{2} \rangle$) exchange their positions in the spectrum and become degenerate when $\epsilon^{'}_{\cal{L}}=0$. At the phase transition point the two  pairs are separated by an energy gap equals to $m_{ \cal{R}}  > 0$ and the ground state consists into the degenerated states $|-\frac{1}{2} \rangle$ and   $|0\rangle_{\epsilon^{'}_{\cal{R}}}$. Hence adding a fermion with spin $\uparrow$ or $\downarrow$
to the system  would costs only the charging energy which scales as $1/L \rightarrow 0$.  Since adding a fermion in the bulk would cost the single particle gap $m$ the zero energy mode is to be localized  
at one of the edges. Yet, from the present Bethe Ansatz analysis, one can not  infer at which of the two edges lies  the zero energy mode, we shall elaborate on this topic in the section  (\ref{Interpretation}).
Similar considerations also hold when considering the other possible phase transitions between phases  $A_{1} \leftrightarrow A_{4}$, $A_{2} \leftrightarrow A_{3}$ and $A_{3} \leftrightarrow A_{4}$.

\subsubsection{Region B\label{regionB}}

As seen from Eq.(\ref{midgap}), when either $|\epsilon^{'}_{\cal{L}}| =1$ or  $|\epsilon^{'}_{\cal{R}}| =1$ one of the two mid-gaps becomes equals to    the single particle gap $m$ itself. When this happens, some of the 
low-lying states (\ref{Astates}) cease to exist and leak into the bulk. As a consequence,   the number of mid-gap states  is reduced when $|\epsilon^{'}_{\cal{L}, \cal{R}}| > 1$. In the region B, either $|\epsilon^{'}_{\cal{L}}| < 1$ and $|\epsilon^{'}_{\cal{R}}| > 1$ or $|\epsilon^{'}_{\cal{L}}| > 1$ and $|\epsilon^{'}_{\cal{R}}| <1$. In this range of twists, there are no boundary string solutions present in the low lying states in contrast to region A. Despite this,
there still exists two  low-lying states with opposite fermion parities, i.e: one with either spins $S^{z}= 1/2$ or  $S^{z}=- 1/2$ and the other with $S^{z}= 0$. We further distinguish between 8 such phases $B_{j=1,...,8}$.
In the phases $(B_1, B_2)$ and  $(B_5, B_6)$, $|\epsilon^{'}_{\cal{L}}| < 1$ and $|\epsilon^{'}_{\cal{R}}| > 1$, while in phases $(B_3, B_4)$ and  $(B_7, B_8)$, $|\epsilon^{'}_{\cal{L}}| > 1$ and $|\epsilon^{'}_{\cal{R}}| <1$.
We list below in the Table (\ref{table2}) the low-energy states, as well as their energies, in each phase.

\begin{table}[h]
\centering
\caption{Total spin and energy in the scaling limit of the low-lying states in the phases $B_j$. In the above
 $E_{-\frac{1}{2}}$ and $E_{+\frac{1}{2}}$ are constants and $m_{ \cal{L}}, m_{ \cal{R}}$ are given by Eq.(\ref{bsenergy}).
 Phases $B_1 (B_5)$ are obtained when $m_{ \cal{L}} >0$ while phases $B_2(B_6)$ are obtained when $m_{ \cal{L}} <0$. Similarly phases $B_3 (B_7)$ are obtained when $m_{ \cal{R}} >0$ while phases $B_4(B_8)$ are  obtained when $m_{ \cal{R}} <0$. When either $m_{ \cal{L}}$ or $m_{ \cal{R}}$  changes its sign there is level crossing and a boundary quantum phase transition occurs when $m_{ \cal{L}(\cal{R})}=0$ between $B_1(B_3)$ and  $B_2(B_4)$ phases, as well as between $B_5(B_7)$ and  $B_6(B_8)$ phases.  }
 \begin{tabular}{ccc}
\hline
\hline
  Phases  & \;\;States &\;\; Energies \\
\hline
 $(B_1, B_2)$&$\{|-\frac{1}{2}\rangle, |0\rangle \}$ &  $(E_{-\frac{1}{2}}, E_{-\frac{1}{2}} +  m_{ \cal{L}})$   \\
 $(B_3, B_4)$ & $\{|+\frac{1}{2}\rangle, |0\rangle  \} $       &  $(E_{+\frac{1}{2}}, E_{+\frac{1}{2}} -  m_{ \cal{R}})$  \\
 $ (B_5, B_6)$ & $\{ |+\frac{1}{2}\rangle, |0\rangle  \} $  & $(E_{+\frac{1}{2}}, E_{+\frac{1}{2}} -  m_{ \cal{L}})$\\
$ (B_7, B_8)$& $\{|-\frac{1}{2}\rangle, |0\rangle \}$& $(E_{-\frac{1}{2}}, E_{-\frac{1}{2}} +  m_{ \cal{R}})$\\
\hline
\hline
\end{tabular}
\label{table2}
\end{table}

We observe that the low-lying states in the two pairs of phases $(B_1, B_2)$ and $ (B_7, B_8)$ have the same
spins, $S^z=-1/2$ and $S^z= 0$,  but differ in that their relative (mid-gap) energy depends only either $m_{ \cal{L}}$ or $m_{ \cal{R}}$ and hence on either the left or right twists $\epsilon^{'}_{\cal{L},\cal{R}}$. 
The same situation occurs for  the two pairs of states in  phases $(B_3, B_4)$ and $ (B_5, B_6)$ which
have spins $S^z= 0, +1/2$.  As for the $A_j$ phases, the different $B_j$ phases are mapped onto each other by the $\mathbb{Z}_2$ group generator (\ref{z2epsilon}):
 $B_1 \leftrightarrow B_5$,  $B_2 \leftrightarrow B_6$, $B_3 \leftrightarrow B_7$ and $B_4 \leftrightarrow B_8$.
 However, unlike as for the $A_j$ phases, the two states in each of the $B_j$ phases {\it do not} span 
 a representation of the $\mathbb{Z}_2$ symmetry (\ref{z2epsilon}). Moreover, unlike in the region A,  
 the $B_j$ phases are not invariant under space parity as upon  exchanging the two boundaries:  $B_1 \leftrightarrow B_8$,  $B_2 \leftrightarrow B_7$, $B_3 \leftrightarrow B_6$ and $B_4 \leftrightarrow B_5$.   Finally we  consider the  boundary phase transitions between the $B_j$ phases. The only possible transitions are within each pairs in (\ref{table2}) when the mid-gap, either  $m_{\cal{L}}$ or $m_{\cal{R}}$, closes. Each time there is level crossing between the corresponding  two spin states and, for the same reasons as for the $A_j$ phases, there is a zero energy mode when  $\epsilon^{'}_{\cal{L}}=0$ or $\epsilon^{'}_{\cal{R}}=0$.
 
\subsubsection{Region C\label{regionC}}

When both $|\epsilon^{'}_{\cal{L},\cal{R}}| > 1$ the ground state is unique and there are no mid-gap states
in contrast with regions $A$ and $B$. Unlike in the trivial region (\ref{trivialR}) the spins of the ground states
may  take different values depending on the signs of $\epsilon^{'}_{\cal{L},\cal{R}}$. We distinguish between
four phases $C_{j=1,...,4}$ in the following domains of twists:
$C_1$ for ($\epsilon^{'}_{\cal{L}} > 1, \epsilon^{'}_{\cal{R}}> 1$), $C_2$ for ($\epsilon^{'}_{\cal{L}} <- 1, \epsilon^{'}_{\cal{R}}> 1$), $C_3$ for  $(\epsilon^{'}_{\cal{L}} <- 1, \epsilon^{'}_{\cal{R}}< -1 $)
and $C_4$ for  ($\epsilon^{'}_{\cal{L}} >1, \epsilon^{'}_{\cal{R}}<- 1$). 
 The  ground-states and their spins are listed below in Table (\ref{table3}).
\begin{table}[h]
\centering
\caption{Total spin of the ground-state states in the phases $C_j$.}
\begin{tabular}{cc}
\hline
\hline
  Phases  & \;\;Ground-state  \\
\hline
 $C_1$&$|-\frac{1}{2}\rangle$  \\
 $C_2$ & $|0\rangle$  \\
 $ C_3$ & $|+\frac{1}{2}\rangle$ \\
$ C_4$& $ |0\rangle$\\
\hline
\hline
\end{tabular}
\label{table3}
\end{table}

\section{Interpreting the Bethe Ansatz Results\label{Interpretation}}

\subsection{Topological Regime}

The hallmark of both $A_j$ and $B_j$ phases is the existence of mid-gap states. Consider for
instance the phases $A_1$  ($m_{\cal{L}} >0$,  $m_{\cal{R}} >0$) and $B_1$ ($m_{\cal{L}} >0$).
There are at least two mid-gap states with  energies $m_{\cal{L}} < m$ and $m_{\cal{R}}<m$
in the phase  $A_1$  and one mid-gap state in the phase $B_1$ with energy $m_{\cal{L}} < m$.
Hence adding a fermion with spin $\uparrow$ to the system at one of the boundaries would costs,
up to the charging energy,  an energy smaller than the single particle gap. We therefore expect that bound
states, localized at the ends of the systems, do exist in each of these two phases. The situation is similar
in all other phases where these mid-gap states occur.
The question that naturally arises  at this point  is what is the nature of these bound-states, particularly when expressed 
in the bare fermions basis. Form the Bethe Ansatz point of view it is a highly non trivial problem that would require the knowledge 
of the wave-functions of the low-lying mid-gap states which is still yet a formidable task.  In the following we shall argue that, under
sensible minimal hypothesis, a nice and consistant picture emerges in which phases $A_j$ and $B_j$ can be understood as spin-states 
bound-states localized  at the ends of the system. In the $A_j$ phases the bound state structure is exhausted by {\it fractional} 
spin-$1/4$ at the two edges exactly as at  the topological fixed point.  In the $B_j$ phases the spin states at the ends 
remain {\it un-fractionalized}, i.e. they are spin-$1/2$  localized at either the left or right edge.

\subsubsection{Fractionalized  Region A}

As mentioned above, the presence of mid-gap states has the consequence that the system is capable
of absorbing added fermions at its edge with an energy cost smaller than the single particle gap.
Consider for definiteness  the $A_1$ phase, with $m_{\cal{L}} >0$ and  $m_{\cal{R}} >0$, and ground-state
$|-1/2\rangle$. Consider first adding to the system a fermion of spin $\uparrow$ at  the left and/or the right boundary by acting with the bare fermion operators $\Psi^{\dagger}_{\uparrow}(x\simeq -L/2)$ and/or  $\Psi^{\dagger}_{\uparrow}(x\simeq +L/2)$ on the ground-state. In the large $L$ limit we expect the following overlaps
\bea
\Psi^{\dagger}_{\uparrow}(x\simeq -\frac{L}{2})  |-\frac{1}{2}\rangle &\rightarrow&  |0\rangle_{\epsilon^{'}_{\cal{L}}},  \nonumber \\\Psi^{\dagger}_{\uparrow}(x\simeq +\frac{L}{2})  |-\frac{1}{2}\rangle &\rightarrow & |0\rangle_{\epsilon^{'}_{\cal{R}}}, \nonumber \\
\Psi^{\dagger}_{\uparrow}(x\simeq -\frac{L}{2}) \Psi^{\dagger}_{\uparrow}(x\simeq +\frac{L}{2})  |-\frac{1}{2}\rangle &\rightarrow&  |+ \frac{1}{2}\rangle.
\eea
These processes would cost, in the $L \rightarrow \infty$ limit,  the energies  $m_{\cal{L}} < m$,  $m_{\cal{R}} < m$ and $m_{\cal{L}} +m_{\cal{R}} < 2m$ respectively. One may,   similarly, further consider adding or removing  a fermion with  spin $\downarrow$  or remove a  fermion with a spin $\uparrow$ at either edges by acting on the states (\ref{Astates}) with the operators 
$\Psi^{\dagger}_{\downarrow}(x\simeq \pm L/2)$ and $\Psi_{\uparrow\downarrow}(x\simeq \pm L/2)$. Up to the charging energy, which is zero in the thermodynamical limit, 
one may then easily convince ourselves that all these processes can be reproduced 
by introducing fermion operators   $a^{\dagger}_{\cal{L}, \cal{R}}$, such that
 $\Psi^{\dagger}_{\uparrow}(x\simeq \pm L/2) \sim a^{\dagger}_{\cal{L}, \cal{R}}$ and 
 $\Psi^{\dagger}_{\downarrow}(x\simeq \pm L/2) \sim a^{}_{\cal{L}, \cal{R}}$,
 which act on the  states  (\ref{Astates}) as
 \be
 \label{zeromode}
a^{\dagger}_{\cal{L}, \cal{R}} |-\frac{1}{2}\rangle \equiv |0\rangle_{\epsilon^{'}_{\cal{L}, \cal{R}}}, \; a^{\dagger}_{\cal{L}}a^{\dagger}_{\cal{L}} |-\frac{1}{2}\rangle \equiv |+\frac{1}{2}\rangle,
\ee
with $a^{}_{\cal{L}, \cal{R}} |-1/2\rangle = 0$. The operators $a^{\dagger}_{\cal{L}, \cal{R}}$ ($a^{}_{\cal{L}, \cal{R}}$) create (destroy) a spin $1/2$ at the left and right boundaries at the cost of the mid-gap energies
$m_{\cal{L}, \cal{R}}$. From the Bethe Ansatz perspective,  acting with $a_{\cal{L}, \cal{R}}$ on the state  $|+1/2\rangle$ is equivalent to adding the boundary strings $\lambda_{\epsilon^{'}_{\cal{L}}}$ and $\lambda_{\epsilon^{'}_{\cal{R}}}$ to the $S^z=+1/2$ state. Hence the fermion operators $a^{}_{\cal{L}, \cal{R}}$ actually correspond to genuine bound-states modes. One may repeat the same arguments for any of the
$A_j$ phases with the same conclusions apart from the fact that the mid-gap energies $m_{\cal{L}, \cal{R}}$
may now also be negative. One may write down the effective low-energy   Hamiltonian and spin operator acting on the states (\ref{Astates}) in terms of the boundary bound-states modes
\bea
\label{Hboundarya}
h_B - E_{-\frac{1}{2}}&=& m_{\cal{L}} a^{\dagger}_{\cal{L}} a^{}_{\cal{L}}  +  m_{\cal{R}} a^{\dagger}_{\cal{R}} a^{}_{\cal{R}} \nonumber \\
S^z &=& \frac{1}{2} (a^{\dagger}_{\cal{L}} a^{}_{\cal{L}}  +  a^{\dagger}_{\cal{R}} a^{}_{\cal{R}} - 1).
\eea
The above Hamiltonian describes all $A_j$ phases in the region $-1< \epsilon^{'}_{\cal{L}, \cal{R}}<1$ in which
the mid-gaps range in the interval $-m <  m_{\cal{L}, \cal{R}} < m$. It  reproduces all possible ground-states and mid-gaps states energies of the states  (\ref{Astates}) in all the  $A_{j=1,...,4}$ phases (see table (\ref{table2}))
as well as the boundary phase transition lines  between them which are  given by $m_{\cal{L}, \cal{R}}=0$.

We shall now  further {\it assume} that the operators $a^{}_{\cal{L}, \cal{R}}$ commute, in the thermodynamical limit, with the Hamiltonian (\ref{Hamiltonian}), i.e:  $[a^{}_{\cal{L}, \cal{R}}, H]=0$. This statement implies   that the four states (\ref{Astates}) generate  four orthogonal towers of excited states that span the whole Hilbert space.  A fact which is consistent with the Bethe Ansatz results. These four  towers are labelled by local, i.e. left and right, fermionic parity quantum numbers $({\cal P}_{\cal{L}}, {\cal P}_{\cal{R}}) = (\pm 1, \pm 1)$ where
 \be
 \label{LRfermionparities}
 {\cal P}_{\cal{L},\cal{R}} \equiv  \sigma^z_{\cal{L},\cal{R}}
=2 a^{\dagger}_{\cal{L},\cal{R}}a^{}_{\cal{L},\cal{R}} -1.
\ee
With these definitions, the total fermion parity  ${\cal P}=(-1)^N= -{\cal P}_{\cal{L}}  {\cal P}_{\cal{R}}$. We list below in the Table \ref{table4} the fermion parities of the four towers of states generated upon the four states (\ref{Astates}).
\begin{table}[h]
\centering
\caption{Local vs total fermionic parities of the low-energy states.}
\begin{tabular}{ccccc}
\hline
\hline
States   & $|-\frac{1}{2}\rangle$ & $ |0\rangle_{\epsilon^{'}_{\cal{L}}}$ & $ |0\rangle_{\epsilon^{'}_{\cal{R}}}$&  $|-\frac{1}{2}\rangle$  \\
\hline
 $ ({\cal P}_{\cal{L}},  {\cal P}_{\cal{R}})  $ & $(+1, +1)$ & $(-1, +1)$ & $(+1, -1)$ & $(-1, -1)$ \\
 ${\cal P}= -{\cal P}_{\cal{L}}  {\cal P}_{\cal{R}}$ & $-1$& $+1$ & $+1$& $-1$ \\
 \hline
\hline
\end{tabular}
\label{table4}
\end{table}  
The above considerations stems from the fractionalization of the $\mathbb{Z}_2$ group (\ref{z2}) between the two edges, i.e:  $\mathbb{Z}_2= \mathbb{Z}_{2,{\cal L}} \otimes \mathbb{Z}_{2,{\cal R}}$ where $\mathbb{Z}_{2,\cal{L},\cal{R}}= \{1, \sigma^x_{\cal{L},\cal{R}}\}$. The generators $\sigma^x_{\cal{L},\cal{R}}$ are defined so that  they reverse the local fermion parity ${\cal P}_{\cal{L},\cal{R}}$ of the states   (\ref{Astates}) and express in  terms of the bound-states modes as  $\sigma^x_{\cal{L},\cal{R}}= (a^{\dagger}_{\cal{L},\cal{R}}+a^{}_{\cal{L},\cal{R}})$. Together with $\sigma^y_{\cal{L},\cal{R}}=-i (a^{\dagger}_{\cal{L},\cal{R}}-a^{}_{\cal{L},\cal{R}})$,
$\sigma^x_{\cal{L},\cal{R}}, \sigma^y_{\cal{L},\cal{R}}$ are the    four Majorana modes, localized at the ends
of the system, associated with the low-energy excitations.  Away from the topological fixed point, i.e. when  $\epsilon^{'}_{\cal{L},\cal{R}} \neq 0$, they are gapped excitations and it is only when 
$\epsilon^{'}_{\cal{L},\cal{R}} \rightarrow 0$ that they become the ZEM of the topological $SSS$-${\widehat{OBC}}$ phase. As wee shall now see, the above fractionalization of the $\mathbb{Z}_2$ symmetry
implies the existence of fractional spin-$1/4$ localized at the two ends of the system.
Indeed, in a system where the total number of particles  $N$ and the total spin $S^z$ are both conserved the total   fermion parity ${\cal P}=e^{-i2\pi S^z}$. We may therefore define {\it fractional} spin-$\frac{1}{4}$ operators $S^z_{\cal{L},\cal{R}}= \frac{1}{4} \sigma^z_{\cal{L},\cal{R}}$ such that ${\cal P}_{\cal{L},\cal{R}} \equiv -i e^{i2\pi S^z_{\cal{L},\cal{R}}}$ and $S^z= S^z_{\cal{L}} + S^z_{\cal{R}}$. These operators act on spin-$\frac{1}{4}$
states localized at the two edges, i.e: $S^z_{\cal{L},\cal{R}}|\pm 1/4\rangle_{\cal{L}, \cal{R}}= \pm1/4 |\pm 1/4\rangle_{\cal{L}, \cal{R}}$,  which span each a representation
of the fractionalized $\mathbb{Z}_{2,\cal{L},\cal{R}}$ groups. We have the correspondence 
\bea
\label{fracspin}
|\pm \frac{1}{2}\rangle &=& |\pm 1/4\rangle_{\cal{L}} \otimes |\pm 1/4\rangle_{\cal{R}},\nonumber \\
|0\rangle_{\epsilon^{'}_{\cal{R}}} &=& |-1/4\rangle_{\cal{L}} \otimes |+1/4\rangle_{\cal{R}}, \nonumber \\
|0\rangle_{\epsilon^{'}_{\cal{L}}} &=& |+1/4\rangle_{\cal{L}} \otimes |-1/4\rangle_{\cal{R}}. 
\eea
In this basis,  we can write the low-energy effective Hamiltonian acting on the boundary states (\ref{Hboundarya})
 in the phase $A_j$ as
\bea
\label{HboundaryA}
h_B - E_{-\frac{1}{2}}&=& \frac{1}{2} (m_{\cal{L}} + m_{\cal{R}}) + { h}_{\cal{L}} S^z_{\cal{L}} +{ h}_{\cal{R}}S^z_{\cal{R}},
\eea
where  ${ h}_{\cal{L}, \cal{R}} = 2 m_{\cal{L}, \cal{R}}$ are effective magnetic fields acting on the localized
spin-$1/4$ operators. Both descriptions (\ref{Hboundarya}) and (\ref{HboundaryA}) are equivalent and valid
in the regime where $-m < m_{\cal{L}, \cal{R}}  < m$. When $|\epsilon'_{\cal{L}}| \rightarrow 1$ or
$|\epsilon'_{\cal{R}}| \rightarrow 1$, $|m_{\cal{L}}| \rightarrow m$ or $|m_{\cal{R}}| \rightarrow m$, some of the bound-states cease to exist because the left boundary term or the right boundary term in the Bethe equations vanishes when $|\epsilon'_{\cal{L}}| \rightarrow 1$ or $|\epsilon'_{\cal{R}}| \rightarrow 1$ respectively. As described in the previous section, when one  of the $|\epsilon_{\cal{L}, \cal{R}}|> 1$ one enters  other phases which low-energy descriptions completely change.

\subsubsection{Un-Fractionalized  Region B}
In the B phases, i.e when  $|\epsilon^{'}_{\cal{L}}| < 1$ and $|\epsilon^{'}_{\cal{R}}| > 1$ or $|\epsilon^{'}_{\cal{L}}| > 1$ and $|\epsilon^{'}_{\cal{R}}| <1$, there   still exists one mid-gap state with energy $m_{\cal{L}}$ or
$m_{\cal{R}}$, respectively. The construction of the low-energy Hamiltonian in these cases proceeds similarly as for the 
A phases though the interpretation of the bound-states modes differs radically. As we shall see, neither the $\mathbb{Z}_2$ group nor the spin fractionalize in these cases. 

Consider first the case where $|\epsilon^{'}_{\cal{L}}| < 1$ and $|\epsilon^{'}_{\cal{R}}| > 1$ that is to say the $(B_1, B_2)$ and $(B_5, B_6)$ phases (see Table (\ref{table2})). As described in (\ref{regionB}) the ground-states in the $B_1$ and $B_5$
phases are $|\pm \frac{1}{2}\rangle$ and have total spins  $\pm \frac{1}{2}$ respectively.
In the Bethe Ansatz approach, they are obtained starting from the reference states with either all spin down or up and  contain no boundary strings nor holes. 

In the $B_1$ ($m_{\cal{L}} >0$) or $B_5$ ($m_{\cal{L}} <0$)  phases one may add a fermion with spin $\uparrow$ or spin $\downarrow$  with the only energy cost of the mid-gap energy  $|m_{\cal{L}}| < m$. Because of the gap
in the bulk, it is energetically more favorable
to add the fermions at one of the edges. Since the energy cost depends only on the left twist $\epsilon^{'}_{\cal{L}}$ we may reasonably assume that  the added, $\uparrow$ and $\downarrow$, fermions are localized at the left edge. We are therefore led to expect the following overlaps:
$\Psi^{\dagger}_{\uparrow}(x\simeq -\frac{L}{2})  |-\frac{1}{2}\rangle \rightarrow  |0\rangle $,
$\Psi^{\dagger}_{\downarrow}(x\simeq -\frac{L}{2})  |+\frac{1}{2}\rangle \rightarrow  |0\rangle$. One may then assume that there exists fermion operators $a^{\dagger}_{\uparrow (\downarrow),\cal{L}}$ that create a bound-states corresponding to an accumulation of  a spin $\pm \frac{1}{2}$ localized at the left edge:
$a^{\dagger}_{\uparrow (\downarrow),\cal{L}}  |(\mp)\frac{1}{2}\rangle=  |0\rangle$. With this definitions  we may  write the Hamiltonian and spin operator for the $(B_1, B_2)$  phases as 
\bea
\label{HB1B2}
h_{\uparrow,\cal{L}}&=& E_{-\frac{1}{2}} + m_{\cal{L}}\;  a^{\dagger}_{\uparrow,\cal{L}}a^{}_{\uparrow,\cal{L}}, \nonumber \\
S^z&=&\frac{1}{2}(a^{\dagger}_{\uparrow, \cal{L}}a^{}_{\uparrow, \cal{L}}-1), 
\eea
and those for the $(B_5, B_6)$ phases
\bea
\label{HB5B6}
h_{\downarrow,\cal{L}}&=& E_{+\frac{1}{2}} - m_{\cal{L}} \; a^{\dagger}_{\downarrow,\cal{L}}a^{}_{\downarrow,\cal{L}}, \nonumber \\
S^z&=& \frac{1}{2}(1-a^{\dagger}_{\downarrow,\cal{L}}a^{}_{\downarrow,\cal{L}}).
\eea
As one varies $-m <m_{\cal{L}}<m$ both Hamiltonians $h_{\uparrow,\cal{L}}$ and $h_{\downarrow,\cal{L}}$
reproduce the mid-gap structure of the phases $(B_1, B_2)$ and $(B_5, B_6)$  as given in Table (\ref{table2}).
The two Hamiltonians (\ref{HB1B2},\ref{HB5B6}) transform into each other under the $\mathbb{Z}_2$ group (\ref{z2},\ref{z2epsilon}): $S^z  \rightarrow - S^z$ and  $h_{\downarrow,\cal{L}}  \rightarrow h_{\uparrow,\cal{L}}$ with   $a^{\dagger}_{\downarrow,\cal{L}} \rightarrow a^{\dagger}_{\uparrow,\cal{L}}, m_{\cal{L}} \rightarrow -  m_{\cal{L}}$. 

Notice that, in the above description,
it is assumed that in the two ground-states of the $B_1$
and $B_5$ phases, i.e: $|\mp\frac{1}{2}\rangle$, the spin is delocalized in the bulk. In this description the
$S^z=0$ states contain a localized bound-state corresponding to an accumulation of spin $\pm \frac{1}{2}$ which counterbalances that of the delocalized spin in the ground-state. An alternative view is possible in which
the states $|\mp\frac{1}{2}\rangle$ are seen to host a bound-state corresponding to an accumulation of spins
$\mp\frac{1}{2}$ localized at the left boundary. In such a description the two $S^z=0$ states are free of localized
spins at the edge. The latter description can be obtained from the above ones  by the transformation: 
$a^{\dagger}_{\downarrow,\cal{L}} \leftrightarrow a^{}_{\uparrow,\cal{L}}$ upon which the Hamiltonians  and spin operators (\ref{HB1B2},\ref{HB5B6})  
become 
\bea
\label{HB1B2prime}
h_{\uparrow,\cal{L}} &\rightarrow& h^{'}_{\downarrow,\cal{L}}= E_0  - m_{\cal{L}} \; a^{\dagger}_{\downarrow,\cal{L}} a^{}_{\downarrow,\cal{L}}, \nonumber \\
S^z &\rightarrow& S^z=-\frac{1}{2} a^{\dagger}_{\downarrow,\cal{L}}a^{}_{\downarrow,\cal{L}}, \eea
and
\bea
\label{HB5B6prime}
h_{\downarrow,\cal{L}} &\rightarrow& h^{'}_{\uparrow,\cal{L}}= E_0  + m_{\cal{L}} \; a^{\dagger}_{\uparrow,\cal{L}} a^{}_{\uparrow,\cal{L}}, \nonumber \\
S^z &\rightarrow& +\frac{1}{2} a^{\dagger}_{\uparrow,\cal{L}}a^{}_{\uparrow,\cal{L}}.
\eea
Deciding which of the two descriptions is the correct one would require the knowledge of the ground-states wave functions in the Bethe Ansatz which, as already mentioned, is a difficult task. In any case, we can still infer that there exist a bound-state
localized at the left boundary corresponding to either
the accumulation of a spin $\pm \frac{1}{2}$ at the left edge.

The case where $|\epsilon^{'}_{\cal{L}}| < 1$ and $|\epsilon^{'}_{\cal{R}}| > 1$ is simply obtained by exchanging the left and right edges $\cal{L} \rightarrow \cal{R}$. The Hamiltonians and spin operators  for the    $(B_3, B_4)$ and $(B_7, B_8)$ phases take the forms of Eqs.(\ref{HB1B2},\ref{HB5B6}), or alternatively Eqs.(\ref{HB1B2prime},\ref{HB5B6prime}), upon changing 
$a^{\dagger}_{\downarrow,\uparrow,\cal{L}} \rightarrow
a^{\dagger}_{\downarrow,\uparrow,\cal{R}}$. For the same reasons as discussed, the bound-states
correspond to an accumulation of spins $\mp\frac{1}{2}$ localized at the right boundary $|\mp\frac{1}{2}\rangle$. 

In each of the $B_j$ phases, when either $|\epsilon^{'}_{\cal{L}}| \rightarrow 1$ or 
$|\epsilon^{'}_{\cal{R}}| \rightarrow 1$, the bound-states
cease to exist and leak into the bulk. When
 the $|\epsilon^{'}_{\cal{L},  \cal{R}}| > 1$ one enters
 one of  the $C_j$ phases where the ground-state is unique and have the same spin quantum numbers as the ground-states in the corresponding $B_j$ phases, i.e: 
 $(B_1, B_8) \rightarrow C_1$ with ground-state $|-\frac{1}{2}\rangle$, $(B_2, B_3) \rightarrow C_2$ with ground-state $|0\rangle$, $(B_4, B_5) \rightarrow C_3$ with ground-state $|+\frac{1}{2}\rangle$ and
 $(B_6, B_7) \rightarrow C_4$ with ground-state $|0\rangle$. 
 
 \section{Discussions}
 
 In this work we have presented   exact results concerning
   1-D superconductors in the presence of integrable boundary fields. The effects of the interactions  between the bulk fermions and the boundaries have proven to lead to a most interesting rich and complex phase diagram. In particular, we showed that the boundary fields can drive the seemingly topologically
   trivial spin-singlet superconductor  toward a topological phase in which the system hosts  zero energy  protected energy Majorana modes at its edges. This topological phase is stabilized by  a $\mathbb{Z}_2$-symmetric topological fixed point  which corresponds to twisted $\widehat{OBC}$ at $\epsilon^{*}_{\cal{L}}=\epsilon^{*}_{\cal{R}}=\pi/2$.
   The above fixed point controls a hole region in its vicinity, the region A in the text,  where the zero energy Majorana modes becomes mid-gap states. In this region, the zero energy modes acquire
 small gaps, $m_{\cal{L}}$ and $m_{\cal{R}}$,  which remain in the mid-gap region of the superconductor. 
 We argued that these bound-states, as at the topological fixed point,  can still be described in terms of fractionalized spins $\frac{1}{4}$ localized at the two edges.  The effect of the $\mathbb{Z}_2$ symmetry breaking boundary fields being captured 
 by magnetic fields acting only at the left and right edges on the spin $\frac{1}{4}$. When one departs to far
 from the topological fixed point, some of the bound-states leave the mid-gap region of the superconductor, leak into the bulk, and  fractionalization is lost. Despite this, there still
 exists a region of boundary fields, coined region B in the text, where one mid-gap state is still present. In this region,  the system still hosts a localized bound-state which corresponds to an accumulation of a spin $\pm \frac{1}{2}$ at either the left or the right boundary. The physics described above is  reminiscent of Andreev bound-states in a superconductor\cite{SenguptaMidgap, NagaiMidgap, colinmidgap}. In this respect,  our exact results, for the present
 1-D charge conserving model,  show that their natures may change as a function of the boundary fields. Finally, for too large departures of the topological fixed point, the bound-states are lost and the ground-state of the system is unique: this is  the region C in the text. Despite this, in this region, the spin of the ground-state {\it still} depends on the boundary fields and may have spins $S^z=- \frac{1}{2}, 0$ or $S^z=+ \frac{1}{2}, 0$. This is to be contrasted with what happens for small twists close to the trivial fixed point corresponding to $OBC$
 where the ground-state, being unique, has $S^z=0$ independently of the twists.
 This calls naturally for the question of how do one interpolates between the two  topological and trivial fixed points.  
 This is a highly non trivial problem in general since one do not expect universal answers away from any fixed point. 
 We though hope to come with some answers in the near future.
  
  So far the results presented in this work are valid in the weak-coupling,
  or strong quantum regime, i.e: $|g_{\parallel, \perp}| << 1$, where
  universal answers can be obtained in the scaling limit. In this respect
  we notice that the exact expressions for the mid-gap energies
bear a remarkable simple universal expression, 
i.e:  $m_{\cal{L}, \cal{R}}= m \sin \frac{\pi}{2} \epsilon^{'}_{\cal{L}, \cal{R}}$,
 in terms of the RG invariants of the problem the superconducting $m$ and
 renormalized twists $\epsilon^{'}_{\cal{L}, \cal{R}}$. 
 This is a highly non trivial result for an interacting fermion problem.
 Preliminary calculations show that   they match with the expressions obtained in both 
 the semi-classical approximation where $g_{\parallel} >> 1$ and  $|g_{\perp}| << 1$, and 
 at the Luther-Emery point \cite{LutherEmery} where $g_{\parallel}= \pi/2$ and  $|g_{\perp}| << 1$.
 In the latter case, where the Hamitonian (\ref{Hamiltonian}) becomes that of free 
 massive spinless fermions, the bound-states structure described in this work may be
 seen as being  of the  Jackiw-Rebbi type \cite{JackiwRebbi, Jackiw}. This gives hope
 that our results may extends to the strong couplings. 
 We plan in the near future to extend our work to the
 massive Thirring model to study the strong coupling physics.
 
 Given the strikings effects of the boundary fields  discussed in this work,  similar phenomena
 are expected to occur in the much more intricate problems involving quantum impurities at the edge which induce dynamical boundary conditions. The simplest  case of a single $S=\frac{1}{2}$ Kondo  impurity at the edge of a  superconductor has been studied recently.
 It was shown that such a system displays screened and  un-screened phases separated by a phase transition \cite{PRA1}.
 One may further show that near the boundary between
 these two phases, mid-gap states are formed with the gap closing at 
 the  quantum phase transition. This will be the subject of a forthcoming publication \cite{ParmeshKondo}.

\acknowledgements
 The authors wish to thank Colin  Rylands for interesting and useful discussions.

\bibliography{refpaper}

\begin{widetext}

\appendix

 \section{Bethe Ansatz}
The $U(1)$ Thirring model is given by the Hamiltonian $H= \int_{-L/2}^{L/2} dx\;   {\cal H}$ where 
\bea \label{Hamiltonian}
 {\cal H}&=&  -i v \left(  \psi^{\dagger}_{Ra} \partial_x \psi_{Ra} -  \psi^{\dagger}_{La} \partial_x \psi_{La}\right)
 \\
&+&
   \psi^{\dagger}_{Ra} \psi_{Rb}  [  \,g_{\parallel}\;  \sigma^z_{ab} \sigma^z_{cd} + g_{\perp}\; (\sigma^x_{ab} \sigma^x_{cd}+ \sigma^y_{ab} \sigma^y_{cd})] \psi^{\dagger}_{Lc}\psi_{Ld}.\nonumber 
\eea
In the above equation,  $\sigma^{x,y,z}$ are the Pauli matrices and  the two-components spinor fields  $\psi_{L(R)}(x)$,
which  describe left and right moving fermions  carrying spin 1/2 with components $a=(\uparrow, \downarrow)$.

We apply the following boundary conditions 
\bea
\Psi_{Ra}(L/2)&=&-B^{\cal{R}}_{ab}\Psi_{Lb}(L/2),\\ \Psi_{Ra}(-L/2)&=&-B^{\cal{L}}_{ab}\Psi_{Lb}(-L/2),
\eea
where
\bea
\label{twist1}
B^{\cal{R}}_{ab}=\frac{1}{\cosh(\frac{f}{2})}\left(\begin{array}{cc} \cosh(\frac{u}{2} (\frac{f}{u}+i\epsilon^{'}_{\cal{R}}))&0\\0&\cosh(\frac{u}{2}(\frac{f}{u}-i\epsilon^{'}_{\cal{R}}))\end{array}\right),
\nonumber \\
\eea

and 

\bea
\nonumber
\small{B^{\cal{L}}_{ab}=\frac{1}{\cosh(\frac{f}{2})}\left(\begin{array}{cc} \cosh(\frac{u}{2}(\frac{f}{u}-i+i\epsilon^{'}_{\cal{L}}))&0\\0&\cosh(\frac{u}{2}(\frac{f}{u}-i-i\epsilon^{'}_{\cal{L}}))\end{array}\right)}.
\eea

These boundary conditions break the space parity $x\leftrightarrow -x$ symmetry and also break the $\mathbb{Z}_2$ symmetry associated with $ \Psi_{L(R),\uparrow} \leftrightarrow \Psi_{L(R),\downarrow}$ transformation.
 Where $f$ and $u$ are parameters related to $g_\parallel$ and $g_\perp$ through the following relations \cite{duty,Japaridze}
  \bea\cos(u)=\frac{\cos(g_{\parallel})}{\cos(g_{\perp})}, \hspace{2mm} \frac{\sin(u)}{\tanh(f)}=\frac{\sin(g_{\parallel})}{\cos(g_{\perp})}.\eea

The parameters $\epsilon'_{\cal{L}}=2\epsilon_{\cal{L}}/u$ and $\epsilon'_{\cal{R}}=2\epsilon_{\cal{R}}/u$ are the asymmetric boundary parameters associated with the left and the right boundaries respectively. 

\subsection{N-particle solution}
\label{sec:AppendixNparticles}

The Hamiltonian commutes with total particle number, $N=\int \psi_+^\dag(x)\psi_+(x)+\psi_-^\dag(x)\psi_-(x)$ and $H$ can be diagonalized by constructing the exact eigenstates in each $N$ sector. 
Since $N$ is a good quantum number we may construct the eigenstates by examining the different $N$ particle sectors separately. We start with $N=1$ wherein we can write the wavefunction as an expansion in plane waves,

\bea\nonumber
\ket{k}=\sum_{a_j=\uparrow\downarrow,\sigma=\pm}\int_{-\frac{L}{2}}^{\frac{L}{2}}\mathrm{d}x\, e^{i\sigma kx} A^\sigma_{a_1}    \psi^{\dagger}_{\sigma,a_1} (x)\ket{0} .
\eea
  $\ket{0}$ is the drained Fermi sea and $A^\sigma_{a_1}$ are the amplitudes for an electron with chirality $\sigma$ and spin $a_1$. The two boundary S-matrices $S^{1R}_{a_1b_1},S^{1L}_{a_1b_1}$ exchange the chirality of a particle. 
\bea A^-_{a_1}=S^{1R}_{a_1b_1} \; A^+_{b_1} \\A^+_{a_1}=S^{1L}_{a_1b_1} \; A^-_{b_1}.\eea

 The asymmetric boundary conditions \eqref{twist1} lead to the following boundary S-matrices \bea S^{1R}_{ab}=B^{R\dagger}_{ab}, \;\; S^{1L}_{ab}=B^{L\dagger}_{ab}. \eea Applying the boundary condition at the left boundary also quantizes the bare particle momentum $k$. We now consider the two particle sector, $N=2$, were the bulk interaction plays a role.
Since the two particle interaction is point-like
we may divide configuration space into regions such that
the interactions only occur at the boundary between two
regions. Therefore away from these boundaries we write
the wave function as a sum over plane waves so that the most general two particle state can be written as
\bea\label{2particle}
\ket{k_1,k_2}&=& \sum_{\sigma,a} \int_{-\frac{L}{2}}^{\frac{L}{2}}\mathrm{d}^2x\,F_{a_1a_2}^{\sigma_1\sigma_2}(x_1,x_2)e^{\sum_{j=1}^2i\sigma_jk_jx_j}\psi^{\dagger}_{\sigma_1a_1}(x_1)\psi^{\dagger}_{\sigma_2a_2}(x_2) \ket{0},
\eea 
where we sum over all possible spin and chirality configurations  and the two particle wavefunction, $F_{a_1a_2}^{\sigma_1\sigma_2}(x_1,x_2)$ is split up according to the ordering of the particles,
\bea
 F_{a_1a_2}^{\sigma_1\sigma_2}=A_{a_1a_2}^{\sigma_1\sigma_2}[12]\theta(x_2-x_1)+A_{a_1a_2}^{\sigma_1\sigma_2}[21]\theta(x_1-x_2).
\eea
The amplitudes $A_{a_1a_2}^{\sigma_1\sigma_2}[Q]$ refer to a certain chirality and spin configuration, specified by $\sigma_j$, $a_j$ as well as an ordering of the particles in configuration space denoted by $Q$. For $Q=12$ particle $1$ is to the left of particle $2$ while for $Q=21$ the  order of the particles are exchanged.
Applying the Hamiltonian to \eqref{2particle} we find that it is an eigenstate with energy $E=k_1+k_2$ provided that these amplitudes are related to each other via application of $S$-matrices. The amplitudes which differ by exchanging the chirality of the leftmost or the rightmost particle are related by the boundary S-matrices.
\bea
A^{\sigma_1-}[12]=S^{2R} \;A^{\sigma_1+}[12], \;\; A^{+\sigma_2}[12]=S^{1L} \;A^{-\sigma_2}[12],\\
A^{-\sigma_2}[21]=S^{1R}\;A^{+\sigma_2}[21],\;\; A^{\sigma_1+}[21]=S^{2L}\;A^{\sigma_1-}[21].\eea
 As discussed above in the one particle case, the boundary S-matrices are  $S^{1R}=B^{R\dagger}, \; S^{1L}=B^{L\dagger},\; S^{2R}=B^{R\dagger}, \; S^{2L}=B^{L\dagger}$. For ease of notation we have suppressed spin indices. It is understood that $S^{1R},S^{1L}$ act in the spin space of particle 1 whereas $S^{2R},S^{2L}$ act in the spin space of particle 2.

There are two types of two particle bulk $S$-matrices denoted by $S^{12}$ and $W^{12}$ which arise due to the bulk interactions and relate amplitudes which have different orderings. The first relates amplitudes which differ by exchanging the order of particles with opposite chirality 
\bea
A^{+-}[21]=S^{12}A^{+-}[12],\\
A^{-+}[12]=S^{12}A^{-+}[21],
\eea
where  $S^{12}$ acts on the spin spaces of particles 1 and 2. Explicitly it is given by, \cite{duty}
\bea\label{S12} S^{ij}= \left(\begin{array}{cccc} 1&&&\\&\frac{\sinh(f)}{\sinh(f+\eta)}&\frac{\sinh(\eta)}{\sinh(f+\eta)}&\\&\frac{\sinh(\eta)}{\sinh(f+\eta)}&\frac{\sinh(f)}{\sinh(f+\eta)}&\\&&&1\end{array}\right).\eea
where $\eta=-iu$ and $f$, $u$ are related to $g_{\parallel}$ and $g_{\perp}$ through the relations $\cos(u)=\frac{\cos(g_{\parallel})}{\cos(g_{\perp})}$ and $\frac{\sin(u)}{\tanh(f)}=\frac{\sin(g_{\parallel})}{\cos(g_{\perp})}.$ In obtaining the above form of the S matrix we have ignored an unimportant overall factor.
Whilst the second {{type of $S$-matrix}} relates amplitudes where  particles of the same chirality are exchanged,
\bea
A^{--}[21]=W^{12}A^{--}[12],\\\label{W12}
A^{++}[12]=W^{12}A^{++}[21].
\eea
 Unlike \eqref{S12}, $W^{12}$ is not fixed by the Hamiltonian but rather by the consistency of the construction. This is expressed through the Yang-Baxter equations 
\bea\label{BYB1}
S^{23}\;S^{13}\;W^{12}&=&W^{12}\;S^{13}\;S^{23},\\
W^{23} \;W^{13} \;W^{12}& = &W^{12} \;W^{13}\; W^{23},\\  S^{2R}\;S^{12}\;S^{1R}\;W^{12}&=&W^{12}\;S^{1R}\;S^{12}\;S^{2R},\\\label{BYB2}S^{2L}\;S^{12}\;S^{1L}\;W^{12}&=&W^{12}\;S^{1L}\;S^{12}\;S^{2L},\eea

which need to be satisfied for the eigenstate to be consistent. We take $W^{12}=P^{12}$ which can be explicitly checked to satisfy \eqref{BYB1}-\eqref{BYB2}. The relations \eqref{2particle}-\eqref{W12} provide a complete set of solutions of the two particle problem.

We can now generalize this to the $N$-particle sector and find that the eigenstates of energy $E=\sum_{j=1}^Nk_j$ are of the form
\bea\label{NparticleS}
\ket{\{k_j\}}=
\sum_{Q,\vec{a},\vec{\sigma}}\int \theta(x_Q) A^{\{\sigma\}}_{\{a\}}[Q] \prod_j^N e^{i\sigma_j k_jx_j}\psi^{\dagger}_{a_j\sigma_j}(x_j)\ket{0}.
\eea
Here we sum over all  spin and chirality configurations specified by $\{a\}=\{a_1\dots a_N\}$, $\{\sigma\}=\{\sigma_1\dots \sigma_N\}$ as well as different orderings of the $N$ particles. These different orderings correspond to elements of the symmetric group $Q\in \mathcal{S}_N$. In addition $\theta(x_Q)$ is the Heaviside function which is nonzero only for that particular ordering. 
As in the $N=2$ sector the amplitudes $A^{\vec{\sigma}}_{\vec{a}}[Q]$ are related to each other by the various $S$-matrices in the same manner as before i.e. amplitudes which differ by changing the chirality of the leftmost particle are related by $S^{jL}$, the amplitudes which differ by changing the chirality of the rightmost particle are related by $S^{jR}$ and the amplitudes which differ by exchanging the order of opposite or same chirality particles are related by $S^{ij}$ and $W^{ij}$ respectively. The consistency of this construction is then guaranteed by virtue of these $S$-matrices satisfying the following Yang-Baxter equations\cite{Sklyannin, Cherednik, ZinnJustin}
\bea\label{YB1}
W^{jk} \;W^{ik}\; W^{ij} &=& W^{ij} \;W^{ik} \;W^{jk},\\ \label{YB2}
S^{jk}\;S^{ik}\;W^{ij} &=& W^{ij}\;S^{ik}\;S^{jk},
\\ S^{jR}\;S^{ij}\;S^{iR}\;W^{ij}&=&W^{ij}\;S^{iR}\;S^{ij}\;S^{jR}\label{YB3},\\S^{jL}\;S^{ij}\;S^{iL}\;W^{ij}&=&W^{ij}\;S^{iL}\;S^{ij}\;S^{jL}\label{YB4},\eea
 Where $W^{ij}=P^{ij}$ and as before the superscripts denote which particles the operators act upon.

\subsection{Bethe equations}
\label{sec:AppendixBE}

In this section we derive the Bethe equations (3). 
Enforcing the boundary condition at $x=-L/2$ on the eigenstate \eqref{NparticleS} we obtain the following eigenvalue problem which constrains the $k_j$,
\bea
e^{-2ik_jL}A^{\{\sigma\}}_{\{a\}}[\mathbb{1}]=\left(Z_j\right)^{\{\sigma\},\{\sigma\}'}_{\{a\},\{a\}'} A^{\{\vec{\sigma}'\}}_{\{\vec{a}'\}}[\mathbb{1}].
\eea
Here $\mathbb{1}$ denotes the identity element of $\mathcal{S}_N$, i.e. $\mathbb{1}=12\dots N$ and the operator $Z_j$ is the transfer matrix for the $j^\text{th}$ particle given by
\bea
Z^j=W^{jj-1}\dots W^{j1} S^{jL}S^{j1}...S^{jj-1}S^{jj+1}...S^{jN}S^{jR}W^{jN}...W^{jj+1}\eea
where the spin indices have been suppressed. This operator takes the $j^\text{th}$ particle from one side of the system to the other and back again, picking up $S$-matrix factors along the way as it moves past the other $N-1$ particles, first as a right mover and then as a left mover.   Using the relations \eqref{YB1}- \eqref{YB4} one can prove that all the transfer matrices commute, $[Z^j,Z^k]=0$ and therefore are simultaneously diagonalizable. In order to determine the spectrum of $H$ we must therefore diagonalize $Z^j,~\forall j$. Here we choose to diagonalize $Z^1$. To do this we use the method of boundary algebraic Bethe Ansatz \cite{Sklyannin, Cherednik, ODBA}.  In order to use this method we need to embed the bare S-matrices in a continuum \cite{trieste} that is, we need to find the matrices $R(\lambda)$, $K(\lambda)$ such that for certain values of the spectral parameter $\lambda$, we obtain the bare S-matrices of our model. Note that the S matrix $S^{12}$ is of the form of $XXZ$ $R$ matrix 

\bea R^{ij}(\lambda)=\left(\begin{array}{cccc} 1&&&\\&\frac{\sinh(\lambda)}{\sinh(\lambda+\eta)}&\frac{\sinh(\eta)}{\sinh(\lambda+\eta)}&\\&\frac{\sinh(\eta)}{\sinh(\lambda+\eta)}&\frac{\sinh(\lambda)}{\sinh(\lambda+\eta)}&\\&&&1\end{array}\right).\eea

 We can see that \small{$R^{ij}(0)=W^{ij}, \hspace{2mm} R^{ij}(f)= S^{ij}$}. The $K$ matrices are given by \cite{Sklyannin} \bea K^{jR}(\lambda)= \frac{1}{\cosh(\lambda)}\left(\begin{array}{cc} \cosh(\lambda-i\epsilon_{\cal{R}})&0\\0&\cosh(\lambda+i\epsilon_{\cal{R}})\end{array}\right),\eea 
 \bea K^{jL}(\lambda)= \frac{1}{\cosh(\lambda)}\left(\begin{array}{cc} \cosh(\lambda+\eta-i\epsilon_{\cal{L}})&0\\0&\cosh(\lambda+\eta+i\epsilon_{\cal{L}})\end{array}\right)\eea

 and are related to the boundary S-matrices as $S^{jR}=K^{jR}(f/2)$, $S^{jL}=K^{jL}(f/2)$.
 The transfer matrix{{ $Z_1$ }}is related to the Monodromy matrix $\Xi_{\tau}(\lambda)$ as $Z^1=t(\frac{f}{2})=\Tr_{\tau} K^{\tau L}(\frac{f}{2})\Xi^{\tau}(\frac{f}{2})$, where
\bea\Xi^{\tau}(\lambda)= R^{1\tau}(\lambda+\frac{f}{2})...R^{N\tau}(\lambda+\frac{f}{2})K^{\tau R}(\lambda)R^{N\tau}(\lambda-\frac{f}{2})...R^{1\tau}(\lambda-\frac{f}{2}).\eea
Here $\tau$ represents {{an  }}auxiliary space and $\Tr_\tau$ represents the trace in the auxiliary space.
{{Using the properties of the $R$ matrices one can prove that $[t(\lambda),t(\mu)]=0$ \cite{ODBA} and by expanding $t(\mu)$ in powers of $\mu$, obtain infinite set of conserved charges which guarantees integrability}}. By following the Boundary Algebraic Bethe Ansatz approach we obtain the following Bethe equations in the region $A$, corresponding to the reference state with all up spins 

\bea\label{EBAE1}
e^{2ik_jL}=\Pi_{i}^{R,L}\beta^{-1}_{\delta i}(f/2)\;\Pi_{\alpha=1}^M\Pi_{\sigma=\pm} \;  \gamma(f/2,\sigma\lambda_\alpha,u/2), \;\; \\ \gamma(x,y,z)=\frac{\sinh(x+y-iz)}{\sinh(x+y+iz)}, \beta_{\delta R}(x)=\frac{\cosh(x-i\delta\epsilon_{\cal{R}})}{\cosh(x)}, \beta_{\delta L}(x)=\frac{\cosh(x-i\delta\epsilon_{\cal{L}})}{\cosh(x)}\eea

 where $\delta=+,-$ corresponds to Bethe reference state with all up spin and all down spin respectively.  $\lambda_\alpha$, $\alpha=1,\dots,M$ are the Bethe roots which satisfy the following equations 
\bea\label{BAE1}
\Pi_{\sigma=\pm,i=R,L}\gamma^N(\lambda_\alpha,\sigma f/2 ,u/2)\gamma(\lambda_\alpha,i\pi/2,-(u-2\delta\epsilon_i)/2)=\Pi_{\beta=1,\sigma=\pm}^{M} \gamma(\lambda_\alpha,\sigma \lambda_\beta,u).
\eea
 
 By rescaling $\lambda_\alpha\rightarrow u\lambda_\alpha$ and applying logarithm we obtain the following Bethe equations in the \textit{trivial phase}  
 \bea\label{logBAE1} \sum_{\sigma=\pm}N\Theta(\lambda_\alpha+\sigma f/2u ,1/2)-\sum_i^{R,L} \Theta(\lambda_\alpha+i\pi/2u,(1-\delta\epsilon^\prime_i)/2)=\sum_{\beta=1}^{M} \sum_{\sigma=\pm}\Theta\left(\lambda_\alpha+\sigma \lambda_\beta,1\right)+2i\pi I_\alpha \eea

\bea\label{logEBAE1}
 k_j=\frac{\pi n_j}{L}+\frac{i}{2L}\left(\sum_i^{R,L}\log[\beta_{\delta i}(f/2)]+\sum_{\beta=1}^M\sum_{\sigma=\pm}\Theta(f/2u+\sigma\lambda_\beta,1/2)\right),
\eea

where $\displaystyle{\Theta(x,y)=\log\left(\frac{\sinh(u(x+iy))}{\sinh(u(x-iy))}\right)}$ and $\epsilon^\prime_i=2\epsilon_i/u$.

\vspace{3mm}

To obtain the Bethe equations corresponding to the \textit{topological phase} where the bulk is in the $STS$ phase with the boundary conditions given by \ref{twist1}, we need to change $g_\perp\rightarrow-g_\perp$. This corresponds to $u\rightarrow-u$, $f\rightarrow-i\pi-f$ \cite{Japaridze}. To keep the sign of $\epsilon'_{\cal{R}}, \epsilon'_{\cal{L}}$ fixed, we need to start with the boundary conditions \eqref{twist1} with $\epsilon_{\cal{R}}\rightarrow-\epsilon_{\cal{R}}$, $\epsilon_{\cal{L}}\rightarrow-\epsilon_{\cal{L}}$. We obtain the following set of Bethe equations

\bea\label{EBAE2}
e^{2ik_jL}=\Pi_{i}^{R,L} \widehat{\beta}_{\delta i}^{-1}(f/2)\;\Pi_{\alpha=1}^M\Pi_{\sigma=\pm} \;  \gamma(f/2,\sigma\lambda_\alpha,u/2), \;\;   \widehat{\beta}_{\delta R}(x)=\frac{\sinh(x-i\delta\epsilon_{\cal{R}})}{\sinh(x)}, \;\; \widehat{\beta}_{\delta L}(x)=\frac{\sinh(x-i\delta\epsilon_{\cal{L}})}{\sinh(x)}
\eea

 \bea\label{BAE2}
\Pi_{\sigma=\pm}\gamma^N(\lambda_\alpha,\sigma f/2 ,u/2) \Pi_i^{R,L} \gamma(\lambda_\alpha,0,-(u-2\delta\epsilon_i)/2)=\Pi_{\beta=1,\sigma=\pm}^{M} \gamma(\lambda_\alpha,\sigma \lambda_\beta,u),
\eea
 Applying logarithm to the above equation and rescaling the Bethe roots we obtain the Bethe equations in phase $2$
\bea\label{logBAE2}
 \sum_{\sigma=\pm}N\Theta(\lambda_\alpha+\sigma f/2u ,1/2)- \sum_i^{R,L} \Theta(\lambda_\alpha,(1-\delta\epsilon^\prime_i)/2) =\sum_{\beta=1}^{M} \sum_{\sigma=\pm}\Theta\left(\lambda_\alpha+\sigma \lambda_\beta,1\right)+2i\pi I_\alpha,
 \eea

\bea\label{logEBAE2}
 k_j=\frac{\pi n_j}{L}+\frac{i}{2L}\left(\sum_i^{R,L}\log[\widehat{\beta_{\delta i}}(f/2)]+\sum_{\beta=1}^M\sum_{\sigma=\pm}\Theta(f/2u+\sigma\lambda_\beta,1/2)\right).
\eea

 Note that when boundary conditions which do not break the $\mathbb{Z}_2$ symmetry are applied the Bethe equations corresponding to the reference state with all up spins are same as those corresponding to the reference state with all down spins. We can obtain the above Bethe equations holding the bulk parameters fixed and by only shifting the boundary parameters $\epsilon_{\cal{R}}\rightarrow \frac{\pi}{2}+\epsilon_{\cal{R}}$, $\epsilon_L\rightarrow \frac{\pi}{2}+\epsilon_{\cal{L}}$. Doing so the bulk remains in the $SSS$ phase but the boundary conditions undergo a non trivial change. Up to an unimportant factor they are given by

 \bea
\label{twist2}
\hat{B}^{\cal{R}}_{ab}=\frac{1}{\sinh(\frac{f}{2})}\left(\begin{array}{cc} \sinh(\frac{u}{2} (\frac{f}{u}+i\epsilon'_{\cal{R}}))&0\\0&-\sinh(\frac{u}{2}(\frac{f}{u}-i\epsilon'_{\cal{R}}))\end{array}\right),
\nonumber \\
\eea

and 

\bea
\nonumber
\hat{B}^{\cal{L}}_{ab}=\frac{1}{\sinh(\frac{f}{2})}\left(\begin{array}{cc} \sinh(\frac{u}{2}(\frac{f}{u}-i+i\epsilon'_{\cal{L}}))&0\\0&-\sinh(\frac{u}{2}(\frac{f}{u}-i-i\epsilon'_{\cal{L}}))\end{array}\right).
\eea

\subsection{Trivial region}
In this section we solve for the distribution of Bethe roots in the ground state in the trivial phase. In the ground state all the Bethe roots take real values. Differentiating \eqref{logBAE2} and noting that $\rho(\lambda)=\frac{d}{d\lambda}\nu(\lambda)$ \cite{trieste}, we obtain the following integral equation, 
\bea
h_{1}(\lambda)&=&\rho_{1}(\lambda)+\sum_{\sigma=\pm}\int_{-\infty}^{+\infty} d\mu \; a_2(\lambda-\sigma\mu)\rho_{1}(\mu),\nonumber \\
\label{deneqn1}
\eea
where $\rho_{1}$ stands for the ground state density distribution in phase $1$ and $h_{1}(\lambda)=Na_1(\lambda+\sigma f/2u)+a_2(\lambda)+ a_1(\lambda)+b_1(\lambda)-b_{(1-\epsilon'_{\cal{R}})}-b_{(1-\epsilon'_{\cal{L}})}$ where
\be
a_n(x)= \frac{u}{\pi} \frac{\sin(n u)}{\cosh(2ux)-\cos(n u)},
\ee
\be
b_n(x)= -\frac{u}{\pi} \frac{\sin(n u)}{\cosh(2ux)+\cos(n u)}.
\ee
Note that we have excluded the root $\lambda=0$ and also applied the restriction $\lambda_\alpha\neq\lambda_\beta$. The above integral equation can be solved by Fourier transformation \cite{Doikoucritical}. We use the following convention \be\label{gsae} \tilde{f}_n(\omega)=\int_{-\infty}^{\infty}e^{i\omega\lambda} f(\lambda), \; f(\lambda)=\frac{1}{2\pi}\int_{-\infty}^{\infty}\hat{f}(\lambda).\ee

We obtain the Fourier transformed density distribution of Bethe roots in the ground state in the trivial phase.

\bea \label{dist1}\tilde{\rho}_1(\omega)=\tilde{\rho}_0(\omega)+\delta \tilde{\rho}_1(\omega)\eea
where
\bea\label{dist0}\tilde{\rho}_0(\omega)=\frac{\left(2N\cos(f\omega/2u)+1\right)+\displaystyle{\frac{1}{\sinh((\pi-u)(\omega/2u))}}\left(\sinh((\pi-2u)(\omega/2u))-\sinh(\omega/2)\right)}{2\cosh(\omega/2)}\eea

\bea \label{dist1}\delta\tilde{\rho}_{tr}(\omega)=\sum_i^{R,L}\frac{\sinh((1-\epsilon^\prime_i)(\omega/2))}{2\sinh((\pi-u)(\omega/2u))\cosh(\omega/2)}.\eea

The number of roots in this ground state is given by \be\label{nroot1} 2M_{tr}+1=\int_{-\infty}^{\infty}d\lambda\rho_1(\lambda). \ee
Using this in \eqref{dist1} and noting that $\tilde{\rho}(0)=\int d\lambda\rho(\lambda)$ we find that the number of roots in the trivial phase in the scaling limit  $u\ll 1$, $\epsilon\ll 1$ and $\epsilon^\prime_i=2\epsilon_i/u$ fixed, is $ M_1=N/2$. We can find the spin of the ground state by using the relation $(S^z)_{tr}=\frac{N}{2}-M_1$. We obtain
\be(S^z)_{tr} = 0.\ee
The above solution is valid for all the values of the parameters $\epsilon'_{\cal{L}}, \epsilon'_{\cal{R}}$. The ground state in this phase is unique, it corresponds to the even parity sector and has total spin $S^z=0$.

\subsection{Topological region}
The system exhibits several different sub-phases within this phase which correspond to different values of the parameters $\epsilon'_{\cal{L}}$ and  $\epsilon'_{\cal{R}}$. In this section we provide the explicit solution for $\epsilon'_{\cal{R}},\epsilon'_{\cal{L}}>0$. 

\subsubsection{Sub phase $A_1$}
This sub-phase corresponds to the values $0<\epsilon'_{\cal{R}}<1$ and $0<\epsilon'_{\cal{L}}<1$. The Bethe equations \eqref{logBAE2} correspond to the Bethe reference state with all up spins. To obtain the ground state we need to consider the Bethe equations corresponding to the Bethe reference state with all down spins, which can be obtained from \eqref{logBAE2} by making the transformation $\epsilon'_{\cal{R}}\rightarrow-\epsilon'_{\cal{R}}, \epsilon'_{\cal{L}}\rightarrow-\epsilon'_{\cal{L}}$. In the ground state all Bethe roots take real values. Let us denote this state by $\ket{-\frac{1}{2}}_{A_1}$. The reason for this notation will become evident soon. All the results from now on will be labeled by the phase they correspond to. These results are presented in the main text, where the labeling corresponding to the phase is suppressed. By following the same procedure as above we obtain the following integral equation 

\bea
h_{-\frac{1}{2},A_1}(\lambda)&=&\rho_{-\frac{1}{2},A_1}(\lambda)+\sum_{\sigma=\pm}\int_{-\infty}^{+\infty} d\mu \; a_2(\lambda-\sigma\mu)\rho_{-\frac{1}{2},A_1}(\mu),\nonumber \\
\label{deneqna11}
\eea

where $\rho_{-\frac{1}{2},A_1}(\lambda)$ corresponds to the density distribution describing the state $\ket{-1/2}_{A_1}$ and $h_{-\frac{1}{2},A_1}=Na_1(\lambda+\sigma f/2u)+a_2(\lambda)+ a_1(\lambda)+b_1(\lambda)-a_{(1+\epsilon'_{\cal{R}})}-a_{(1+\epsilon'_{\cal{L}})}$. 

The above integral equation can be solved by applying Fourier transform. We obtain the following distribution of Bethe roots
\bea \label{dista111}\tilde{\rho}_{-\frac{1}{2},A_1}(\omega)=\tilde{\rho}_0(\omega)+\delta\tilde{\rho}_{-\frac{1}{2},A_1}(\omega)\eea where
\bea \label{dista112}\delta\tilde{\rho}_{-\frac{1}{2},A_1}(\omega)=\sum_i^{R,L}\frac{\sinh((\pi-u(1+\epsilon^\prime_i))(\omega/2u))}{2\sinh((\pi-u)(\omega/2u))\cosh(\omega/2)}.\eea
 
 The number of Bethe roots can be found by using the relation 
 \bea 
\label{nroota11}2M^b_{-\frac{1}{2},A_1}+1=\int_{-\infty}^{+\infty} d\lambda\; \rho^b_{-\frac{1}{2},A_1}(\lambda),
\eea

 from which the $z$-component of spin $(S^z)_{-\frac{1}{2},A_1}$ of the ground state in this subphase is obtained using the relation $S^z_{-\frac{1}{2},A_1}=N/2-M_{-\frac{1}{2},A_1}$. Taking into account that  $\tilde\rho(0)=\int\mathrm{d}\lambda\, \rho(\lambda)$ along with \eqref{dista111} we find that
 \bea
 (S^z)_{-\frac{1}{2},A_1}=-\left(\frac{\pi}{2(\pi-u)}-\frac{u(\epsilon'_{\cal{R}}+\epsilon'_{\cal{L}})}{4(\pi-u)}\right) 
 \eea
In the scaling limit, i.e. when $|g_\parallel |\ll1,|g_\perp|\ll1, u\ll1$ we also need to take $\epsilon_{\cal{R}}\ll1, \epsilon_{\cal{L}}\ll1$ while $\epsilon'_{\cal{R}}=\frac{2\epsilon_{\cal{R}}}{u}, \epsilon'_{\cal{L}}=\frac{2\epsilon_{\cal{L}}}{u}$ are held fixed. We obtain
 \bea 
 (S^z)_{-\frac{1}{2},A_1}=-\frac{1}{2}.
 \eea 

To confirm that we found the ground state, we need to look at other possible solutions to the Bethe equations in this sub-phase $A_1$. It can be shown through counting argument that even though there exists boundary string solutions in the Bethe equations corresponding to Bethe reference state with all down spins, we cannot add them to the above obtained state unless we add holes in the bulk. Hence the above obtained state has the lowest energy compared to all other states that are solutions to the Bethe equations corresponding to Bethe reference state with all down spins. Now we consider the Bethe equations corresponding to Bethe reference state with all up spins \eqref{BAE2}.

Following the same procedure as above we obtain the following integral equation corresponding to the state $\ket{\frac{1}{2}}_{A_1}$ with all real Bethe roots
\bea
h_{\frac{1}{2},A_1}(\lambda)&=&\rho_{\frac{1}{2},A_1}(\lambda)+\sum_{\sigma=\pm}\int_{-\infty}^{+\infty} d\mu \; a_2(\lambda-\sigma\mu)\rho_{\frac{1}{2},A_1}(\mu),\nonumber \\
\label{deneqna12}
\eea
  
where $\rho_{\frac{1}{2},A_1}(\lambda)$ is the density distribution corresponding to the state $\ket{\frac{1}{2}}_{A_1}$ and $h_{\frac{1}{2},A_1}=Na_1(\lambda+\sigma f/2u)+a_2(\lambda)+ a_1(\lambda)+b_1(\lambda)-a_{(1-\epsilon'_{\cal{R}})}-a_{(1-\epsilon'_{\cal{L}})}.$ By applying Fourier transform we obtain

\bea \label{dista12}\tilde{\rho}_{\frac{1}{2},A_1}(\omega)=\tilde{\rho}_0(\omega)-\delta\tilde{\rho}_{\frac{1}{2},A_1}(\omega)\eea where

\bea \label{dista12}\delta\tilde{\rho}_{\frac{1}{2},A_1}(\omega)=\sum_i^{R,L}\frac{\sinh((\pi-u(1-\epsilon^\prime_i))(\omega/2u))}{2\sinh((\pi-u)(\omega/2u))\cosh(\omega/2)}.\eea

The total spin of this state can be obtained by using a relation same as \eqref{nroota11}, we obtain
\bea
 (S^z)_{\frac{1}{2},A_1}=\frac{\pi}{2(\pi-u)}+\frac{u(\epsilon'_{\cal{R}}+\epsilon'_{\cal{L}})}{4(\pi-u)}. 
 \eea

Taking the scaling limit we get
 \bea 
 (S^z)_{\frac{1}{2},A_1}=\frac{1}{2}.
 \eea

By observation one can see that there exists boundary string solutions $\lambda_{\epsilon'_{\cal{R}}}=\pm\frac{i}{2}(1-\epsilon'_{\cal{R}})$,  $\lambda_{\epsilon'_{\cal{L}}}=\pm\frac{i}{2}(1-\epsilon'_{\cal{L}})$ to the Bethe equations \eqref{BAE2}. These boundary string solutions which correspond to boundary bound states can be added to the state $\ket{\frac{1}{2}}_{A_1}$. 
Adding the boundary string $\lambda_{\epsilon'_{\cal{R}}}$ to the state $\ket{\frac{1}{2}}_{A_1}$ we obtain the state $\ket{0}_{\epsilon'_{\cal{R}}}$ who's density distribution $\rho_{\epsilon'_{\cal{R}},A_1}(\lambda)$ satisfies the following integral equation

\bea
h_{\epsilon'_{\cal{R}},A_1}(\lambda)&=&\rho_{\epsilon'_{\cal{R}},A_1}(\lambda)+\sum_{\sigma=\pm}\int_{-\infty}^{+\infty} d\mu \; a_2(\lambda-\sigma\mu)\rho_{\epsilon'_{\cal{R}},A_1}(\mu),\nonumber \\
\label{deneqna13}
\eea

where $h_{\epsilon'_{\cal{R}},A_1}=h_{\epsilon'_{\cal{R}},A_1}-(a_{(1+\epsilon'_{\cal{R}})}(\lambda)+a_{(3-\epsilon'_{\cal{R}})}(\lambda))$. Taking Fourier transform we obtain 

\bea \tilde{\rho}_{\epsilon'_{\cal{R}},A_1}(\omega)=\tilde{\rho}_{\frac{1}{2},A_1}(\omega)+\Delta\tilde{\rho}_{\epsilon'_{\cal{R}}}(\omega), \;\; \Delta\tilde{\rho}_{\epsilon'_{\cal{R}}}(\omega)=-\frac{\sinh((\pi-2u)(\omega/2u))\cosh((1-\epsilon'_{\cal{R}})(\omega/2))}{\sinh((\pi-u)(\omega/2u))\cosh(\omega/2)}.\eea

The number of Bethe roots in this state can be found by the following relation
\bea
\label{nroota12}2M^b_{\epsilon'_{\cal{R}},A_1}-1=\int_{-\infty}^{+\infty} d\lambda\; \rho^b_{\epsilon'_{\cal{R}},A_1}(\lambda),
\eea
from which the $z$-component of spin $(S^z)_{\epsilon'_{\cal{R}},A_1}$ can be obtained using the relation $S^z_{\epsilon'_{\cal{R}},A_1}=N/2-M_{\epsilon'_{\cal{R}},A_1}$. We get
\bea
 (S^z)_{\epsilon'_{\cal{R}},A_1}=\frac{u(\epsilon'_{\cal{R}}+\epsilon'_{\cal{L}})}{4(\pi-u)}. 
 \eea

Taking the scaling limit we get
 \bea 
 (S^z)_{\epsilon'_{\cal{R}},A_1}=0.
 \eea 
 
 Similarly, adding the boundary string $\lambda_{\epsilon'_{\cal{L}}}$ we obtain the density distribution describing the state $\ket{0}_{\epsilon'_{\cal{L}}}$ 
 
 \bea \tilde{\rho}_{\epsilon'_{\cal{L}},A_1}(\omega)=\tilde{\rho}_{\epsilon'_{\cal{L}},A_1}(\omega)+\Delta\tilde{\rho}_{\epsilon'_{\cal{L}}}(\omega), \;\; \Delta\tilde{\rho}_{\epsilon'_{\cal{L}}}(\omega)=-\frac{\sinh((\pi-2u)(\omega/2u))\cosh((1-\epsilon'_{\cal{L}})(\omega/2))}{\sinh((\pi-u)(\omega/2u))\cosh(\omega/2)}.\eea
 
 This state has the same $S^z$ as that of the state obtained above
 
 \bea
 (S^z)_{\epsilon'_{\cal{L}},A_1}=\frac{u(\epsilon'_{\cal{R}}+\epsilon'_{\cal{L}})}{4(\pi-u)}. 
 \eea

Taking the scaling limit we get
 \bea 
 (S^z)_{\epsilon'_{\cal{L}},A_1}=0.
 \eea 

We have obtained three states $\ket{\frac{1}{2}}_{A_1}, \ket{0}_{\epsilon'_{\cal{R}}}, \ket{0}_{\epsilon'_{\cal{L}}}$, given by the distributions 
$\tilde{\rho}_{\frac{1}{2},A_1}(\lambda),\tilde{\rho}_{\epsilon'_{\cal{R}},A_1}(\lambda),\tilde{\rho}_{\epsilon'_{\cal{R}},A_1}(\lambda)$ as solutions to the Bethe equations 
 corresponding to the Bethe reference state with all up spins. We now show that all these states are higher in energy compared to the state $\ket{-\frac{1}{2}}_{A_1}$.

 The energy of a state is given by $E=\sum_{j=1}^N k_j$. By using this, \eqref{logEBAE2} can be expressed as

\bea\label{ena}
E=\sum_j\frac{\pi n_j}{L}+\frac{iD}{2}\left(\sum_i^{R,L}\log[\widehat{\beta_{\delta i}}(f/2)]+\int_{-\infty}^{\infty}d\lambda\rho(\lambda)\Theta(f/2u-\lambda_\beta,1/2)\right).
\eea
where $D=N/L$ is cutoff in the system. The first term is the charge contribution and the second term within the bracket is the spin contribution to the total energy.  
  
Consider the states $\ket{\frac{1}{2}}_{A_1}$, $\ket{-\frac{1}{2}}_{A_1}$. The difference in the energy of these states is 
\bea
E_{\ket{\frac{1}{2}}_{A_1}}-E_{\ket{-\frac{1}{2}}_{A_1}}=\frac{iD}{2}\left( \sum_{i}^{R,L}\log\left[\frac{\widehat{\beta}_{+i}}{\widehat{\beta}_{-i}}\right]+\int_{-\infty}^{\infty}d\lambda(\rho_{\frac{1}{2},A_1}(\lambda)-\rho_{-\frac{1}{2},A_1}(\lambda))\Theta(f/2u-\lambda_\beta,1/2)\right)
\eea

Using Fourier transform the above equation can be written as

\bea
E_{\ket{\frac{1}{2}}_{A_1}}-E_{\ket{-\frac{1}{2}}_{A_1}}=\frac{iD}{2}\left( \sum_{i}^{R,L}\log\left[\frac{\widehat{\beta}_{+i}}{\widehat{\beta}_{-i}}\right]+\int_{-\infty}^{\infty}d\omega\frac{e^{i\omega f/2u}}{\omega} (\tilde{\rho}_{\uparrow\uparrow}(\omega)-\tilde{\rho}_{\downarrow\downarrow}(\omega))\right)
\eea
By using \eqref{dista111},\eqref{dista12} in the above equation and evaluating the integral we get

\bea E_{\ket{\frac{1}{2}}_{A_1}}-E_{\ket{-\frac{1}{2}}_{A_1}}=\frac{iD}{2} \sum_{i}^{R,L}\left(\log\left[\frac{\widehat{\beta}_{+i}}{\widehat{\beta}_{-i}}\right]-\log\left(\frac{\sinh(f/2-i\epsilon'_iu/4)}{\sinh(f/2+i\epsilon'_iu/4)}\right)+\log\left(\frac{\tanh(\pi f/4+i\pi\epsilon'_i/4)}{\tanh(\pi f/4-i\pi\epsilon'_i/4)}\right)\right). \eea

The first two terms cancel each other. After some simplification we obtain

\bea \label{entwo} E_{\ket{\frac{1}{2}}_{A_1}}-E_{\ket{-\frac{1}{2}}_{A_1}}= m\sin(\epsilon'_{\cal{L}}\pi/2)+m\sin(\epsilon'_{\cal{R}}\pi/2).\eea

We now calculate the energy difference between the states $\ket{\frac{1}{2}}_{A_1}$, $\ket{0}_{\epsilon'_{\cal{R}}}$. As seen above, the addition of the boundary string $\lambda_{\epsilon'_{\cal{R}}}$ to the state with spin $S^z=1/2$ leads to a new state with spin $S^z=0$ that includes a boundary excitation. The energy difference between these states up to the chemical potential is given by 
\begin{eqnarray}
\label{ebk}E_B=E_{N}-\frac{1}{2}(E_{N-1}+E_{N+1}).
\end{eqnarray}
Here $E_{N}$ refers to the energy of the state with odd number of particles which, in our system, corresponds to the state with spin $S^z=\pm 1/2$. Similarly $E_{N+1}$ and  $E_{N-1}$ refer to the energies of the states with an {\it even} number of particles and spin $S^z=0$. The latter states include the added boundary string. The expression \eqref{ebk} is defined in \cite{Keselman2018} as the binding energy, which precisely measures the energy cost of adding an electron to the system. 

Using \eqref{ena} in \eqref{ebk}, we find that $E_B$ has two contributions, one from the charge degrees of freedom 
 and one from the spin degrees of freedom: 
 $E_B=E_{\text{charge}}+E_{\text{spin}}$. The charge contribution is given by the charging energy
 \begin{eqnarray}
  E_{\text{charge}}=\sum_{j=1}^{N} \frac{\pi}{L}n_j-\frac{1}{2}\left(\sum_{j=1}^{N+1} \frac{\pi}{L}n_j+\sum_{j=1}^{N-1} \frac{\pi}{L}n_j\right). \nonumber \\
  \label{totenc}
 \end{eqnarray}
 Note that the the charge quantum numbers take all the values from the cutoff $-DL$ upwards. In the ground state with $S^z=\pm 1/2$ they fill all the slots from $n_j=-N \; \text{to}\; n_j=-1$. In the state with one extra particle they fill all the slots from $n_j=-N \;\text{to}\; n_j=0$. In the state with one less particle there is an unfilled slot at $n_j=-1$ which corresponds to a holon excitation. Hence we obtain
 \bea \label{bounden}E_{\text{charge}}=-\frac{\pi}{2L}.\eea

 The spin contribution is given by the expression
 \bea
 E_{\text{spin}}=E_{0}+\int_{-\infty}^{\infty}d\lambda(\rho_{\frac{1}{2},A_1}(\lambda)-\rho_{\epsilon'_{{\cal{R}},A_1}}(\lambda))\Theta(f/2u-\lambda_\beta,1/2), \nonumber \\
 \label{enbsr}
 \eea
 where \bea E_{0}=\frac{iD}{2}\left(\log\left(\frac{\sinh(u(f/2u+i/2(2-\epsilon'_{\cal{R}})))}{\sinh(u(f/2u-i/2(2-\epsilon'_{\cal{R}})))}\right)+\log\left(\frac{\sinh(u(f/2u+i\epsilon'_{\cal{R}}))}{\sinh(u(f/2u-i\epsilon'_{\cal{R}}))}\right)\right).\eea
 
  Evaluating \eqref{enbsr} we find that the spin part of the energy difference between these states is given by \bea  E_{\text{spin}}=-m\sin(\epsilon'_{\cal{R}}\pi/2).\eea 
  
Hence from \eqref{entwo} this state has energy $m_{\cal{L}}=\sin(\epsilon'_{\cal{L}}\pi/2)$ above the state $\ket{-\frac{1}{2}}_{A_1}$.  Similarly we find that the state obtained by adding the boundary string  $\lambda_{\epsilon'_{\cal{L}}}$ has energy  $m_{\cal{R}}=m\sin(\epsilon'_{\cal{R}}\pi/2)$ above the state $\ket{-\frac{1}{2}}_{A_1}$. Hence we have shown that the state $\ket{-\frac{1}{2}}_{A_1}$ described by the root distribution $\rho_{-\frac{1}{2},A_1}(\lambda)$ is indeed the ground state.

In the sub-phases $A_j$, the states with spin $S^z=\pm 1/2$ are obtained from Bethe reference states with all up and all down spins and contain all real Bethe roots. In the sub-phase $A_3$, the two singlet states $\ket{0}_{\epsilon^{'}_{\cal{L}}}$, $\ket{0}_{\epsilon^{'}_{\cal{R}}}$ can be obtained by adding the boundary strings $\lambda_{\epsilon^{'}_{\cal{L}}}$, $\lambda_{\epsilon^{'}_{\cal{R}}}$ respectively to the state with spin $S^z=-1/2$. In the sub-phase $A_2$, the boundary strings can be added to the state with spin $S^z=1/2$ for $|\epsilon^{'}_{\cal{R}}|>|\epsilon^{'}_{\cal{L}}|$ and for $S^z=-1/2$ for $|\epsilon^{'}_{\cal{R}}|<|\epsilon^{'}_{\cal{L}}|$. In the sub-phase $A_4$, the boundary strings can be added to the state with spin $S^z=-1/2$ for $|\epsilon^{'}_{\cal{R}}|>|\epsilon^{'}_{\cal{L}}|$ and for $S^z=1/2$ for $|\epsilon^{'}_{\cal{R}}|<|\epsilon^{'}_{\cal{L}}|$.

\subsubsection{Sub-phase $B_1$}
This sub-phase corresponds to $\epsilon'_{\cal{R}}>1$ and $0<\epsilon'_{\cal{L}}<1$. Same as in the previous section, to obtain the ground state we need to consider the Bethe equations corresponding to Bethe reference state with all down spins. The ground state consists of all real Bethe roots. Following the same procedure as above we obtain the following distribution of Bethe roots

\bea\label{distb11}\tilde{\rho}_{-\frac{1}{2},B_1}(\omega)=\tilde{\rho}_0(\omega)-\delta\tilde{\rho}_{-\frac{1}{2},B_1}(\omega)\eea

where
\bea \label{distb12}\delta\tilde{\rho}_{-\frac{1}{2},B_1}(\omega)=-\sum_i^{R,L}\frac{\sinh((\pi-u(1+\epsilon^\prime_i))(\omega/2u))}{2\sinh((\pi-u)(\omega/2u))\cosh(\omega/2)}.\eea

The number of Bethe roots is given by a relation similar to \eqref{nrootsa11}, using which we obtain for the total spin 
\bea S^z_{-\frac{1}{2},B_1}=\frac{\pi}{2(\pi-u)}-\frac{u(\epsilon'_{\cal{R}}+\epsilon'_{\cal{L}})}{4(\pi-u)}.\eea

Taking the scaling limit we obtain $S^z_{-\frac{1}{2},B_1}=-\frac{1}{2}$. Let us denote this state by $\ket{-\frac{1}{2}}_{B_1}$. 
Similarly to the sub-phase $A_1$, even though there exists boundary string solution it only exists in the presence of holes in the bulk. Hence, the lowest energy state corresponding to the Bethe reference state with all spin down is given by the above distribution. We now consider Bethe equations corresponding to the Bethe reference state with all spin up. Following the same procedure as above we obtain the following distribution 

\bea\label{distb11}\tilde{\rho}_{0,B_1}(\omega)=\tilde{\rho}_0(\omega)-\delta\tilde{\rho}_{0,B_1}(\omega)\eea

where
\bea \label{distb12}\delta\tilde{\rho}_{0,B_1}(\omega)=-\frac{\sinh((\pi-u(1-\epsilon^\prime_{\cal{L}}))(\omega/2u))-\sinh((\pi-u(\epsilon^\prime_{\cal{R}}-1))(\omega/2u))}{2\sinh((\pi-u)(\omega/2u))\cosh(\omega/2)}.\eea

From which we obtain 

\bea S^z_{0,B_1}=\frac{u(\epsilon'_{\cal{R}}+\epsilon'_{\cal{L}})}{4(\pi-u)},\eea

which in the scaling limit gives $S^z_{0,B_1}=0$. Let us denote this state by $\ket{0}_{B_1}$

Similar to the case of all spin down Bethe reference state, any solution with a boundary string requires adding holes in the bulk. Hence the lowest energy state corresponding to the Bethe reference state with all up spin is given by the above distribution. Following the same procedure as in the previous section we can calculate the energy difference between the two states $\ket{0}_{B_1},\ket{-\frac{1}{2}}_{B_1}$ obtained above given by the distributions $\tilde{\rho}_{0,B_1}(\omega), \tilde{\rho}_{-\frac{1}{2},B_1}(\omega)$ respectively. We find that the charge part of the energy is given by \eqref{bounden} and the spin part of the energy is given by

\bea \label{entwob} E_{\ket{0}_{B_1}}-E_{\ket{-\frac{1}{2}}_{B_1}}= m\sin(\epsilon'_{\cal{L}}\pi/2).\eea

Hence we find that the state $\ket{-\frac{1}{2}}_{B_1}$ given by the distribution $\delta\tilde{\rho}_{-\frac{1}{2},B_1}(\omega)$ is the ground state. Following the same procedure as above we can find the low lying states in all the sub-phases $B_j$, where there are two low lying states. In sub-phases $C_j$, there is only one low lying state with spin $S^z=0$ or $S^z=\pm 1/2$ and contains all real Bethe roots.

All the sub-phases and the associated states and their energies are summarized in the figure below.
 
\begin{figure}[ht]
\includegraphics[scale=1]{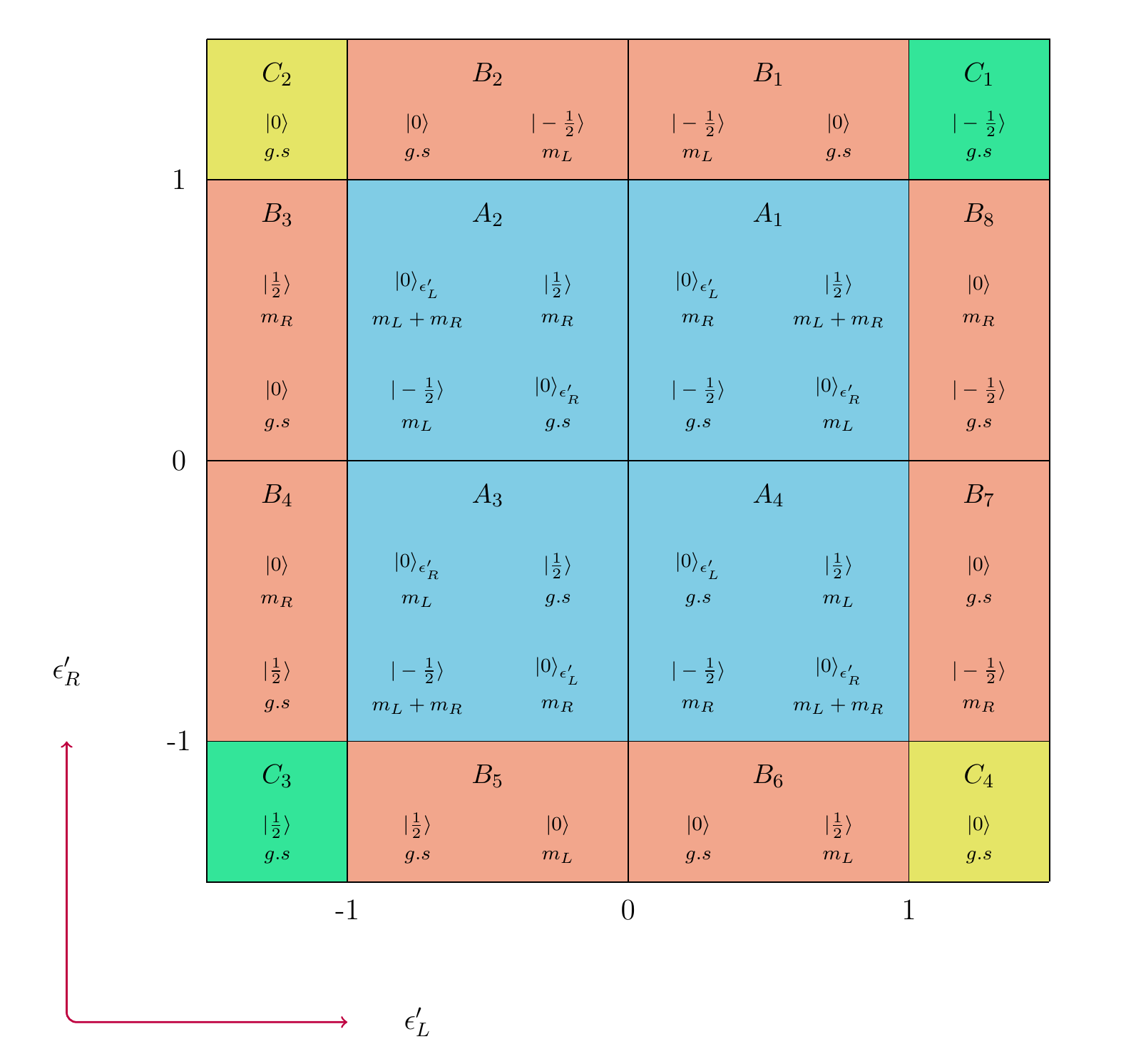}
\caption{In the sub-phases $A_1,A_2,A_3,A_4$ spin fractionalization occurs. In the rest of the sub-phases spin fractionalization does not occur. The ground state is denoted by $g.s$ and the mid-gap energies are denoted by $'m_{\cal{R}}','m_{\cal{L}}'$ where $m_{\cal{R}}=m\sin(\epsilon'_{\cal{R}}\pi/2)$ and $m_{\cal{L}}=m\sin(\epsilon'_{\cal{L}}\pi/2)$. The subscripts on the states which denote the phase are suppressed.}
\end{figure}

\end{widetext}
\end{document}